\RequirePackage{etoolbox}
\csdef{input@path}%
{%
}
\documentclass{elsevierbook}

\usepackage{natbib}
\newcommand{\arcsec}[1]{\textit{"}}

\newcommand\psj{{PSJ}}
\newcommand\apj{{ApJ}}
\newcommand\apjs{{ApJS}}
\newcommand\aj{{AJ}}
\newcommand\mnras{{MNRAS}}
\newcommand\aap{{A\&A}}
\newcommand\pasp{{PASP}}
\newcommand\icarus{{Icarus}}
\newcommand\nat{{Nature}}
\newcommand\apjl{{ApJL}}

\usepackage{hyperref}
\usepackage{color}
\usepackage{bm}
\usepackage{eqnarray}
\usepackage{amsmath} 

\begin{document}

\Frontmatter

\Mainmatter
  
\begin{frontmatter}
\setcounter{chapter}{8}
\chapter{Detecting Moving Objects With Machine Learning}\label{chapter7}


Author: Wesley C. Fraser (Herzberg Astronomy and Astrophysics Research Centre)
\begin{abstract}
 The scientific study of the Solar System's minor bodies ultimately starts with a search for those bodies. This chapter presents a review of the use of machine learning techniques to find moving objects, both natural and artificial, in astronomical imagery. After a short review of the classical non-machine learning techniques that are historically used, I review the relatively nascent machine learning literature, which can broadly be summarized into three categories: streak detection, detection of moving point sources in image sequences, and detection of moving sources in shift and stack searches. In most cases, convolutional neural networks are utilized, which is the obvious choice given the imagery nature of the inputs. In this chapter I present two example networks: a Residual Network I designed which is in use in various shift and stack searches, and a convolutional neural network that was designed for prediction of source brightnesses and their uncertainties in those same shift-stacks. In discussion of the literature and example networks, I discuss various pitfalls with the use of machine learning techniques, including a discussion on the important issue of overfitting. I discuss various pitfall associated with the use of machine learning techniques, and what I consider best practices to follow in the application of machine learning to a new problem, including methods for the creation of robust training sets, validation, and training to avoid overfitting. 
\end{abstract}

\begin{keywords}
\kwd{minor body}
\kwd{moving object}
\kwd{convolutional neural network}
\kwd{machine learning}
\kwd{photometry}
\kwd{digital tracking}
\end{keywords}

\end{frontmatter}

\newcommand{\red}[1]{\textcolor{red}{#1}}



\section{Introduction} \label{intro}
The study of natural moving objects such as the asteroids or comets  has a fruitful history, which has enabled significant insights into the physical and chemical structure of the early protoplanetary disk, the formation processes responsible for the growth of planetesimals and planets, and the delivery of water, organics, and other materials important for the formation of life on the Earth, to name a few topics. The discussion of these topics is beyond the scope of this chapter, but we point the novice reader to review texts such as Asteroids IV \citep{asteroidsiv}, the Trans-Neptunian Solar System \citep{prialnik2019}, Protostars and Planets VII \citep{ppvii}, and the chapters of the upcoming Comets III which are avialable on the arxiv.\footnote{\url{https://arxiv.org/}} The focus of this chapter is the -- often arduous -- task of searching for new minor bodies, a requisite first step in the study of these populations, either as a bulk population, or individually. In this chapter, I first summarize various flavours of classic search techniques which have enabled the current research into minor bodies. I then move on to discussing the nascent field of utilizing machine learning to assist in the search for minor bodies. These techniques are sometimes employed to assist with the most difficult steps in the search process, or to enable new techniques entirely. Hereafter when the distinction matters, I  refer to natural Solar System objects as minor bodies to distinguish them from artificial satellites and other spacecraft. 

The later half of this chapter is dedicated to a discussion of the use of a convolutional neural network (CNN) that I developed to perform source classification in a circumstance that classically requires significant human effort to perform. The CNN I introduce was relatively straight forward to develop, and has been used quite successfully in numerous recent surveys for minor bodies. I use this as an easy to understand example of machine learning (ML), and to discuss some of the pitfalls and best practices in developing an ML tool.


\section{Introduction to the detection of moving objects in astronomical imagery \label{sec:intromops}}
In this section I first introduce classic, non-ML based pipelines for the initial detection of new minor bodies. Generally, the process comes in two stages: detection of a source in astronomical imagery and confirmation of its motion (Section~\ref{sec:search}); and then linking -- connecting different detections of a common object -- together to form arcs from which orbital information and future ephemerides can be gleaned (Section~\ref{sec:link}). I summarize common practices for each part in turn in the following sections. In Section~\ref{sec:shiftnstack} I consider requisite changes to the general moving object procedure required to implement digital tracking techniques, which I refer to as shift'n'stack. In the following section I only discuss recent or ongoing surveys for moving bodies. These were chosen as the most modern examples of search efforts, and only discussed as is necessary to understand the process of common search processes. There is a long and beautiful history of highly successful surveys prior to these which are not summarized here. Rather we point the interested reader to the review texts mentioned above, as well as references in the papers discussed in this section.

\subsection{The Image Search \label{sec:search}}
Conceptually, the search for moving bodies is quite simple: one must simply find sources in a frame, and identify those that are not stationary. This simple statement hides a long history of search efforts of various scales and difficulties. I make no attempt to fully summarize this topic, which would warrant a dedicated textbook, but rather, just point out some common techniques used in the field.

The simplest search technique is to acquire multiple temporally nearby images of a field and search for sources that move. For example, the Asteroid Terrestrial-impact Last Alert System (ATLAS) is an on-going survey which typically utilizes four 30~s exposures aimed at a common point on the sky and that span a 1-hour interval \citep{Tonry2018}. This image cadence allows the detection of motion of minor bodies with rates of motion at least $\sim0.5 \mbox{ \arcsec/hr}$, or roughly half the width of a point-source in the ATLAS imagery. Those imagery undergo image differencing from a background sky map to remove stationary sources. Over the short 1-hour baseline of the imagery, the majority of minor bodies exhibit nearly linear motion enabling the relatively straight forward step of building up \emph{tracklets}, or groupings of nearby detections that show motion compatible with expectations of bound moving minor bodies. 

The above paragraph provides a feel for the general steps taken in a search for moving bodies:

\begin{itemize}
    \item Acquire coincident imagery of a field of interest, with a baseline and cadence that enables detection of motion of the objects of interest. The  imagery can span temporal baselines as short as a few minutes for asteroids which exhibit angular rates of motion at opposition of many tens of \arcsec/hour, or as long as a few hours to detect the more distant Kuiper Belt population which exhibit opposition rates of motion of only a few \arcsec/hour.\footnote{The opposition angular rate of motion of a minor body on a circular orbit and with zero ecliptic inclination and at heliocentric distance $r$ (geocentric distance $\Delta=r-1$) in au is  approximately given by $\dot{\theta} \sim 148 \left[\frac{1}{\Delta} - \frac{1}{r^{3/2}}\right]$~\arcsec/hr.} 
    \item Sort candidate non-stationary sources from stationary sources. This is most robustly done using image differencing \citep[e.g.,][]{Alard1998} to remove background galaxies and stars, but can also simply be done by spatially associating overlapping detections revealing objects that do not move. This leaves behind \emph{transient} sources that are often, but not always moving bodies. 
    \item Find groups of sources that are compatible with the range of motions that could be exhibited by the minor bodies of interest. As hinted at above, this is most commonly linear motion, and often restricted to motions that would only be exhibited by objects with bound orbits, e.g., bound to the Sun for minor bodies, or bound to the Earth for artificial satellites. These groups are commonly referred to as \emph{tracklets} and represent likely real minor bodies. 
\end{itemize}

\noindent
The detections that are assembled into tracklets can be contaminated by a variety of sources, including cosmic rays, detector and processing defects, subtraction residuals, real astronomical stationary sources (e.g., supernovae), and just plane old noise spikes in the data. This is often the stage at which sources need vetting by a human operator, and can represent a huge portion of the human effort required of a project.\footnote{vetting is a right of passage in some groups}

I point the interested reader to a few important recent surveys. ATLAS has been mentioned already, which represents a state of the art in surveys for asteroids, near-Earth asteroids (NEAs) and potentially hazardous asteroids (PHAs). The ATLAS moving object pipeline is largely based on the code developed for, and lessons learned by Pan-STARRS (A. Fitzsimmons, personal communication). Pan-STARRS, or the Panoramic Survey Telescope and Rapid Response System \citep[][]{Chambers2016} is an on-going survey with sensitivity to asteroids as well as more distant -- and hence fainter bodies such as Kuiper Belt Objects (KBOs) and Centaurs. The notable aspect of this survey is that its pipeline is capable of building tracklets of observations that span a few days over which the trajectories exhibited by minor bodies are non-linear \citep{Denneau2013}. The Outer Solar System Origins Survey \citep[OSSOS; ][]{Bannister2018} was specifically designed to observe faint and slow moving KBOs and other distant bodies. OSSOS utilized triplets of imagery acquired in a single night that span a few hours with which to find slow moving objects. Candidate triplets were generated using the routines of \citet{Bernstein2000} and every candidate detection was manually vetted by human operators, thereby forgoing the need for image subtraction. This is just a small selection of recent surveys and by no means fully describes this deep and active field of research.

\subsection{The Linking Stage \label{sec:link}}
The process of connecting tracklets of a common object is commonly referred to as \emph{linking}. Linking is useful in numerous ways. It naturally extends the temporal length of the arc for the body, thereby making any measure of its orbital parameters more accurate. If the arc-length is long enough, it tends to secure the body against future loss -- where the ephemeris uncertainty for the object becomes so large it can no longer be distrinugished from other nearby moving bodies, thus necessitating a full rediscovery. Most importantly for the discussion here, linking can provide some measure of good:bad vetting of tracklets. In much the same way that multiple individual detections grouped together into a tracklet provide veracity to the detections being that of a real moving object, if tracklets can be linked together into an orbit that is not unphysical, that link provides some veracity to the detections and tracklets together. As this can be done in a mostly automated sense, linking can in some circumstances reduce the human cost burden.

In concept, the process of linking is quite simple: a set of tracklets or other candidate observations of the same source are passed to an orbit fitting routine, which returns a set of orbital parameters and uncertainties (often referred to as \emph{the orbit}), and a measure of goodness of fit, which is usually a chi-squared, or a root-mean-square (RMS) measure of the astrometric residuals, or likelihood value. It is then up to the user to decide if the orbit fit quality is high enough (RMS is low enough) to be considered ``good''. Orbit fitting software packages that are commonly used are BKorbit \citep{Bernstein2000}, Orbfit\footnote{\url{http://adams.dm.unipi.it/orbfit/}}, or the more modern HelioLinc \citep{Holman2018}. The exact process varies from survey to survey.

The OSSOS survey mixes detection and linking together, whereby an orbit fit to a tracklet of a triplet of discovery observations is propagated to observations taken within $\sim5$~days of discovery, and candidate moving sources are searched in a singlet image. A new track and orbit fit spanning 5 days is created, and propagated to observations roughly a month from discovery, and then to opposition observation years after discovery. This procedure builds up orbits of exceptional quality and reliability, but comes at high human cost as every stage of linking requires human confirmation.

The Pan-STARRS objects are built up in a conceptually similar manner to OSSOS. Starting from a quad set of images with about 15 minutes of spacing between each image, a tracklet is  made. Those tracklets that could not already be associated with known moving bodies are searched for linkages to quad tracklets of other nights. After a certain number of tracklets are linked, the candidate is designated a real object. Additional linkages are searched for in all prior observations taken by the Pan-STARRS survey. The advantage of this technique over the OSSOS method is in the reduced human burden - at no point in the linking process are individual images considered for possible new links, but rather only tracklets are considered (usually four detections in a tracklet) where the likelihood of the tracklet being that of a real object is high. The advantage of the OSSOS pipeline over the Pan-STARRS method is a decrease in the requisite telescope time required to produce reliable orbits of newly detected objects, but comes at increased human cost.  

The process of orbit linking and fitting is somewhat of an art, and we encourage the interested user to explore the references provided.

\subsection{Digital Tracking Techniques \label{sec:shiftnstack}}
\begin{figure}[h]
\includegraphics[width=0.67\textwidth]{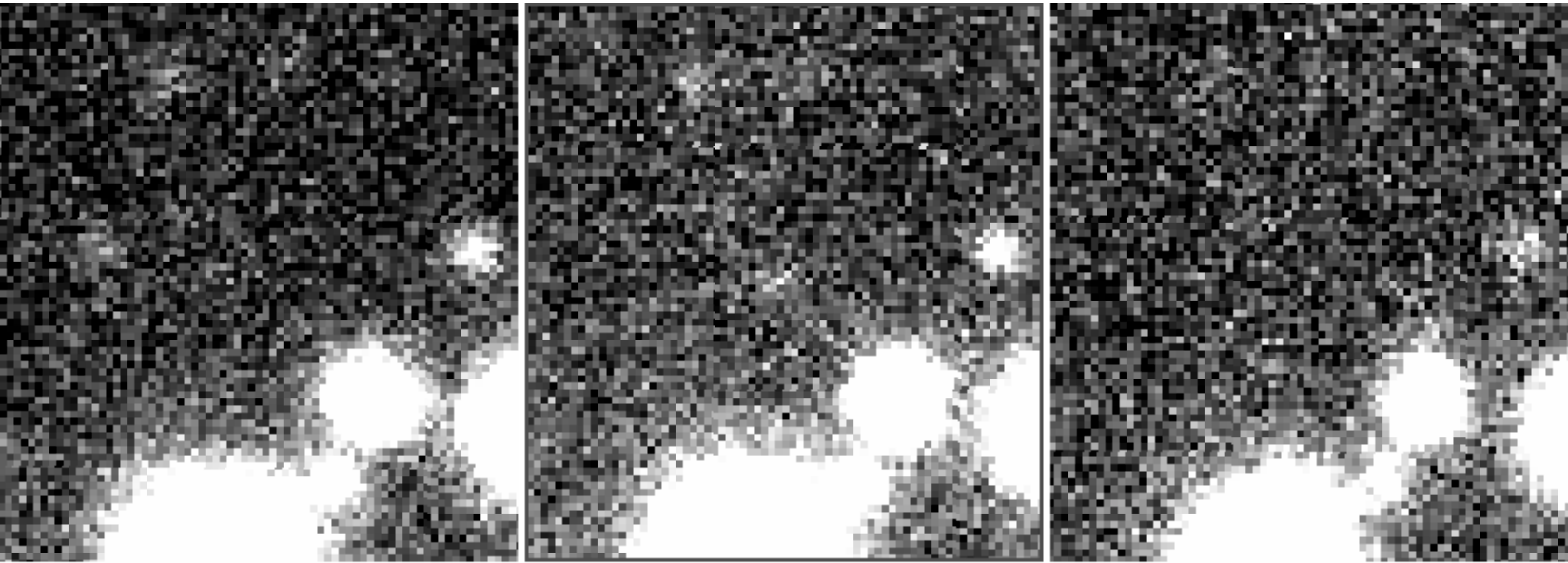}
\includegraphics[width=0.95\textwidth]{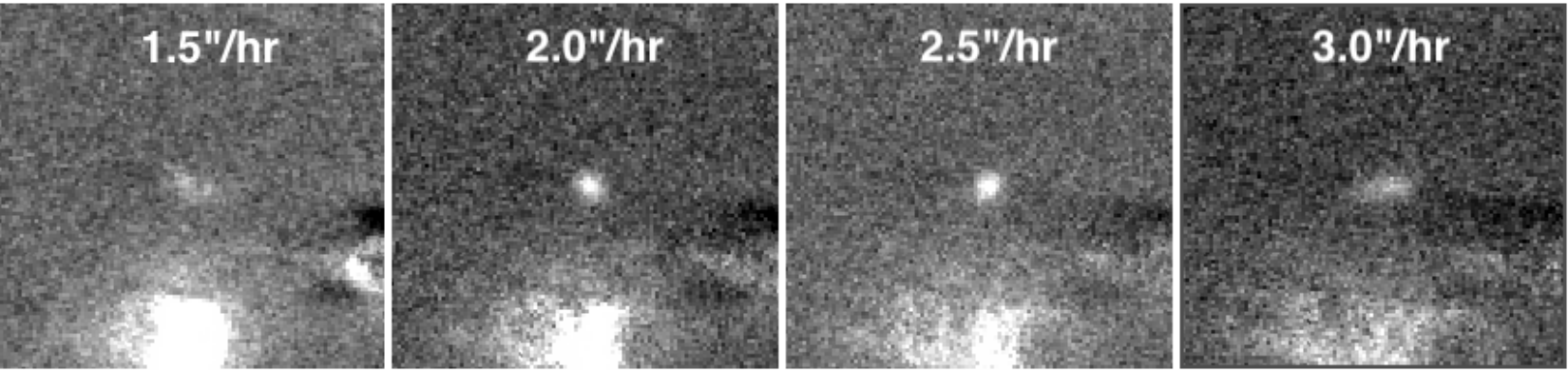}
\caption{\textbf{Top:} Three images centred on a real Kuiper Belt Object discovered as part of the New Horizons Search for distant Kuiper Belt Objects \citep{Fraserlpsc2023}. The source is just below the noise floor of the images. \textbf{Bottom:} The shift'n'stack sequence of $\sim100$ images, shifted at four different linear rates of motion of the same source shown above. The source's true rate of motion is 2.4~"/hr, and when the shift rate is near that value, the source becomes an easy detection. \label{fig:sns}}
\end{figure}

Digital tracking, or shift'n'stack is a process by which temporally nearby images are shifted and stacked to search for fainter sources than would otherwise be visible in a single image. That is, an object's motion through a series of images can be counteracted through an appropriate backwards shift, allowing all of the source's flux contained in those images to be stacked in a single coincident point source (see Figure~\ref{fig:sns}). This process directly enables a much deeper search for moving bodies with limiting magnitudes that often approach the limiting magnitude of a single long exposure of the same time as the image sequence.  This search process comes with a high computational burden, as image stacks need to be produced at every possible trajectory, and for every possible region where the objects may reside (normally the entire field of view).

Moving object detection has a long history of finding ultra-faint sources. This is especially true in the effort to find Kuiper Belt Objects \citep{Luu1998, Gladman2001, Bernstein2004, Fraser2008, Fuentes2009} and other distant, slow moving bodies \citep{Kavelaars2004,Holman2004, Ashton2020}. For these populations, angular rates of motion are highly linear for the duration of a night, and so the shift'n'stack routine can be charaterized by only two dimensions -- rate of motion in Right Ascension and Declination, or equivalently rate and angle. Applications to more proximate populations are less common due to the higher apparent angular accelerations they exhibit which make their apparent trajectories non-linear much more rapidly, though see \citet{Heinze2015}. Once the trajectories of objects become non-linear, either due to proximity, or due to length of image sequence, the computational burden is dramatically increased. For this reason, non-linear shift'n'stack remains an unsolved problem and active area of research.

Beyond the massively increased computational expense, shift'n'stack historically has required significant human vetting effort, primarily as a result of a massive ramp up of false sources driven by noise spikes that show up at low signal-to-noise, as well as other types of false positive sources that do not plague non-stacking search techniques. A particularly nefarious example is from faint diffraction spikes associated with bright stars. As an image sequence is acquired, those diffraction spikes rotate around the star. A linear shift of those images can cause the spikes from different images to overlap in an e stack, which produces a somewhat round source that can slip past many filters implemented to reject false positives. This and other sources of false positive generally necessitate the manual vetting of each and every candidate, resulting in human cost that is much higher than non-stacking techniques. I discuss the use of ML to tackle this particular problem in Section~\ref{sec:ML-sns}.

In the general, shift'n'stack routines replace the discovery-to-tracklet steps. A shift'n'stack detection provides both a position, and a rate of motion. This information replaces the tracklet, and is sufficient to search for a set of shift'n'stack discoveries taken at a different epoch for objects that match the orbit predicted from the detection in the first epoch. This is the general search approach that is being used for the shift'n'stack component of the the DECam Ecliptic Exploration Project \citep[DEEP;][]{Trilling2023,Napier2023}, an on-going search for new targets for observations with the New Horizons Spacecraft \citep[][and also Fraser et al. in prep]{Fraserlpsc2023} and the Classical and Large-A Solar System Survey \citep[CLASSY;][]{Fraser2023acm}. Each of the four fields of DEEP will receive observations at opposition and one month after opposition in the discovery year, and then be re-observed one, and two years after discovery. From those four visits, observations of common sources will be linked and orbits constructed \citep{Trujillo2023}. CLASSY uses an approach inspired by the OSSOS discovery and linking technique: a field is visited 3 times on three separate nights near opposition and within $\sim5$~nights. Each visit consists of 3 total hours of integration time spanning at least 4~hours. Each of those visits is searched with shift'n'stack techniques and tracklets of candidate sources across all three nights are created. Each of these tracklets is propagated to shift'n'stack search epochs at plus and minus one month of discovery to provide confirmation of real objects and to enable initial orbit fits.

Through the proliferation of graphics processor units (GPUs) and some clever programming, the compute burden to perform the shift'n'stack part of a search has been greatly reduced. By virtue of their design, GPUs can make short work of the median or mean image stacking process reducing the computational time by orders of magnitude. The Kernel Based Moving Object Detection \citep[KBMOD][]{Whidden2019,Smotherman2021} software package provides generalized linear shift'n'stack search capabilities. 
As well \citet{burdanov2023} present a similar tool for digital tracking.

There are other less utilized search techniques that hold some resemblance to recent ML based search efforts. I highlight a clever effort by \citet{Fuentes2010} who searched for trailed sources in long-exposure observations from the Hubble Space Telescope (HST). Due to the motion around the Earth, most minor bodies exhibit noticeable parallax as witnessed by the HST resulting in non-linear trailed arcs for Solar System bodies arcs with characteristic motions that reflect the orbit of the HST. In their search, they generated candidate templates of possible trails that might be exhibited by distant KBOs. By template matching, they found a handful of faint and particularly distant objects that went previously undetected. 

\section{Applications of Machine Learning \label{sec:ML-applications}}
In this section I summarize a select set of past and current ML efforts designed to find moving objects\footnote{While I try to be thorough and complete, inevitably some papers will be missed.}. By nature of the fact that the majority of moving sources are discovered in imagery, most modern ML-based applications of moving object detection are based on CNNs of some flavour or other. Though we start out with a likely more familiar ML technique in Section~\ref{sec:clustering}.

\subsection{Clustering in Moving Object Detection \label{sec:clustering}}
Some of the earliest uses of ML in moving object detection are not with neural networks or decision trees which are what is commonly thought of when the term \emph{machine learning} is mentioned. Early search efforts have made use of clustering algorithms to assist in moving object detection. For example, HelioLinc mentioned above requires clustering of detections to sort real moving objects from the chaffe. In the original implementation, a KD-tree \citep{Kubica2007} with euclidean distance on velocity and distance parameters is used. The KBMOD routine also uses the Density-Based Spatial Clustering of Applications with Noise (DBSCAN) algorithm in position and velocity to remove the candidate shift'n'stack sources that are most likely to be false from the final returned list.

Clustering algorithms in general require selection of parameters that determine the output. To achieve best results,  these \emph{hyper parameters} are tuned to maximize performance according to some metric relevant to the circumstance in which the clustering algorithms are used. This is akin to the hyper parameter selection and training that is required of what most readers would consider modern machine learning, albeit usually with simpler training efforts and often more easily understood metrics. Similarly, much like each clustering algorithm is just that -- an algorithm, so is any CNN or similarly complex ML technique, and the choices in network architecture are analogous to choice in clustering algorithm. 

I bring up clustering as a familiar and comfortable concept that many readers may have already used that is also essentially a machine learning technique, in hopes to assuage some of the unreasonable skepticism and fear some have over the uses of ML.

\subsection{Moving Object Detection and Classification in Direct Images}
    
\begin{figure}[h]
\includegraphics[width=0.95\textwidth]{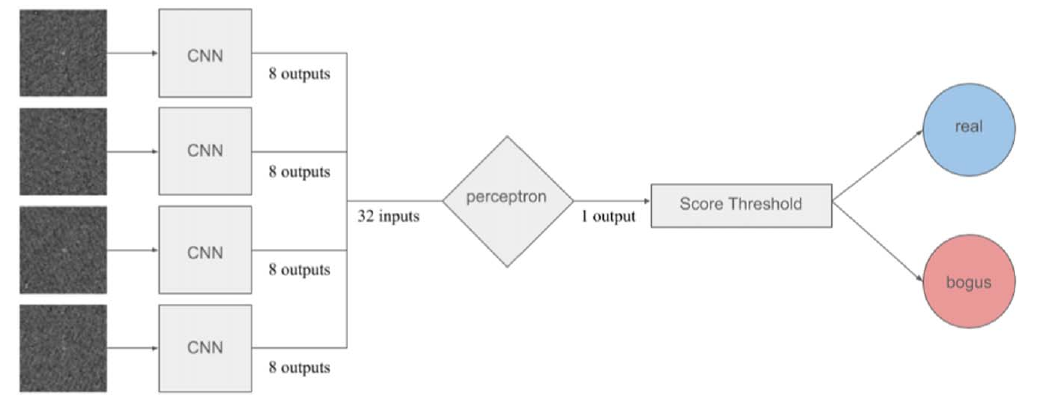}
\caption{The custom architecture utilized by \citet{ChybaRabeendran2021} where each image of an ATLAS quad is passed through a pretrained ResNet-18 (labelled as ``CNN''), and the confidence outputs are concatenated into a single 32-element vector that is passed through the classifier percepteron. Reproduced from Figure~6 of that work. \label{fig:atlas-ml}}
\end{figure}

\begin{figure}[h]
\includegraphics[width=0.95\textwidth]{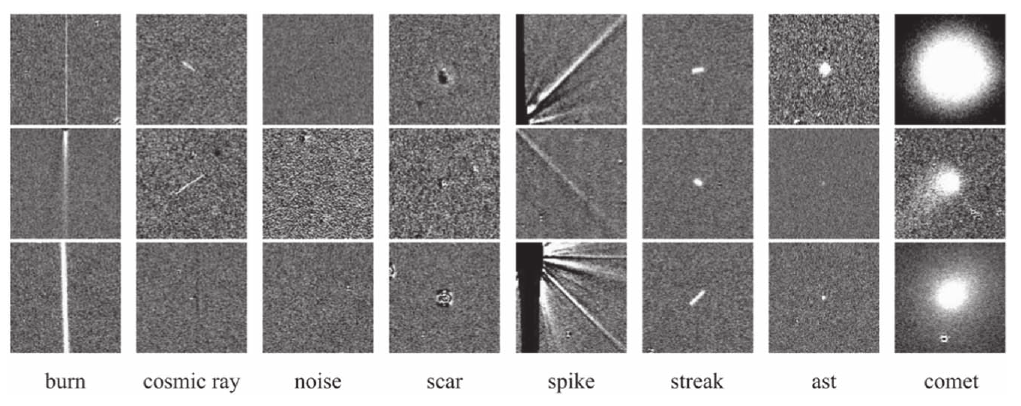}
\caption{Examples of the 8 different classes utilized in the classification search of \citet{ChybaRabeendran2021}. The three real minor body classes are shown at right, with the 5 bogus classes resulting from detector and pipeline artifacts. Reproduced from Figure~3 of that work. \label{fig:atlas-classes}}
\end{figure}

I first discuss the work of \citet{ChybaRabeendran2021} who apply a ResNet-based custom neural network to discovering asteroids in the ALTAS survey (see the introductory chapter for an introduction to the ResNet). Recall that the classic moving object search utilized by ATLAS makes use of a quad set of coincident and temporally close exposures. The authors adopt a novel  architecture to perform binary classification -- does or does not contain a moving body -- in image quads that were previously identified as containing four separate images of a candidate real moving object (see Figure~\ref{fig:atlas-ml}). Specifically, the network makes use of the pretrained ResNet-18 model that feeds a pair of dense layers. Through this structure each image of the quad is passed, with the outputs consisting of an 8 element vector with values $\mathbb{R} \in [0,1)$. Those values represent confidences of the source at the centre of the image belonging to one of eight different classes: five bogus classes including diffraction spikes, detector and pipeline defects; and three real classes including fast moving \emph{streak} objects, point-like slow moving objects, and comets (see Figure~\ref{fig:atlas-classes}). 

The novelty of their work rests with the percepteron classifier network they develop, which takes as input a 32 element confidence vector which is the concatenated confidences of the four image outputs of the ResNet-18. The percepteron returns a likelihood that the image quad contains one of the three types of real moving object. 

The authors trained only the percepteron stage of their custom network, preserving the weights of the ResNet-18. They focused on maximal discovery. That is, they trained to ensure that real minor bodies were labelled as such as often as possible, without focus on the precision or recall performance, which are more commonly maximized. This resulted in a moderately low 15\% false positive rate, and a tiny 0.4\% false negative rate. Importantly, the network resulted in a 90\% reduction in the final human vetting time required to declare a source as real. 

The main weakness of this work was the tiny training sample they made use of. Data from only one lunation was used in training, with only $\sim500$ sources per each of the eight classes. In some sense it is surprising that they have such a positive result, though much of this success is likely driven by the perfectly labelled training set they utilized, and the fact that they only trained the weights of the percepteron. While it is likely that inclusion of more data would enable a significant reduction it the false positive rate, a main takeaway from this work for the aspiring ML user is that good results can be found even with moderately small, but \emph{high quality} training sets.

The results of \citet{ChybaRabeendran2021}, reaffirm the common thought that CNNs can provide highly reliable type classifications for sources. For example, that network does a satisfactory job of identifying cometary sources in ATLAS image quads. Another effort to identify comets in astronomical imagery is presented in \citet{Duev2021}. In that effort, they utilize a fairly advanced network structure to confirm the presence of comets in image cutouts of Zwicky Transient Facility (ZTF) data that was already flagged as possibly containing a comet. There routine utilizes networks that output segmentation maps to perform binary classification as to whether or not an image contains a comet, as well as perform regression on the location of the candidate comet in those images. 

\begin{figure}[h]
\includegraphics[width=0.95\textwidth]{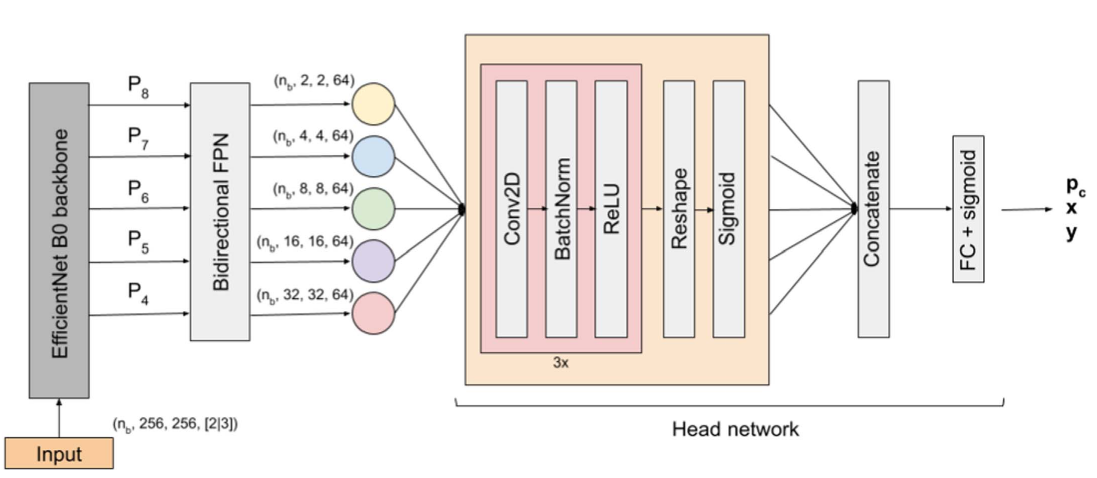}
\caption{The bi-FPN based network used by \citep{Duev2021} to identify comets in ZTF imagery. The network consists of two parts: an EfficientNet-B0 and biFPN section that creates multi-scale feature maps from the input imagery, and a CNN+percepteron head network that performs the binary classification and regression on comet position. Reproduced from Figure~3 of that work. \label{fig:bifpn}}
\end{figure}

\citet{Duev2021} make use of an EfficientNet (version B0) and bi-directional feature pyramid network  \citep[bi-FPN;][]{Tan2020} to create a multi-scale segmentation map. A simple way to think of the segmentation map is a pixel map of the input image that contains large values where pixels in the original image likely contain a comet, and small values where they likely do not. The bi-FPN can result in a segmentation map that preserves morphological information about a source at multiple scales - one might immediately recognize the important of this for comets which show a broad range of morphologies. This segmentation map is fed to a so-called head network which is essentially a standard shallow CNN+percepteron that outputs a vector $[p_c, x, y]$ where $p_c$ is the probability of an image containing a comet, and x and y are the predicted locations.  Their network is presented in Figure~\ref{fig:bifpn}.

Their network takes as input a three-layer image. The first layer is a 256x256 pixel cutout that has previously been identified as plausibly containing a comet. The second layer is a background template of the same region generated from other epochs of ZTF data, and the third layer is the image subtraction of the background template from the first image.

The training was supervised. The training data they utilize reflects the rarity of real comets, with a prelabeled training set that is highly unbalanced, with 5x more cutouts \emph{not} containing comets than cutouts that do. To counteract this, they make use of sample weights to produce a network that can successfully identify images containing a comet. Without sample weights, the network would only have the ability to identify images that \emph{do not} have comets. We discuss the importance of balanced training data further below. Positive labelled training samples were generated using ephemerides of known comets tracked in the Jet Propulsion Laboratory Horizons Service\footnote{\url{https://ssd.jpl.nasa.gov/}} with manual labeling. Other negative labelled imagery was included (though it seems possible that a few of those images may actually contain an unknown comet). The training data also included comets identified with a simple ResNet-based classifier. The resulting training set size was 22,000 samples (compare this with the smaller sample size of \citet{ChybaRabeendran2021} discussed above). 

This impressive work resulted in an enviable false negative rate of $\sim1.7\%$ with a rough position error on predicted comet location of only a few pixels. The proof is in the pudding; they report the discovery of comet C/2020 T2 (Palomar)

\subsection{Trailed Moving Source Detection \label{sec:ML-trailed}}

The majority of effort by the community towards developing machine learning algorithms for detection of minor bodies has been put into searching for streaks of moving objects within  long exposures. These telltale signs are quite common to astronomical surveys that have focus beyond the detection of minor bodies, and so do not take care to avoid circumstances where objects might trail. While not nominal for detection, trailed minor bodies have a unique morphology that is not produced by any other real astronomical source, thereby affording use of some machine learning techniques that are designed to be sensitive to differences in source morphology.

\citet{Jeffries2023} make use of a so-called U-Net \citep{Ronneberger2015} which they train to produce segmentation maps of regions of imagery likely to contain streaks of moving objects, both natural and artificial. The authors fully simulated an astronomical dataset containing streaks, point sources, and noise, and thus also simulated a perfectly labelled training dataset. They made use of a pretrained U-net encoder stage, and only trained the decoder stage. 

The authors found that their streak-detecting U-Net was able to detect and locate streaks with a similar performance to state of the art non--ML-based algorithms, but with a $10-20\times$ higher throughput, though their results did come with some other minor weaknesses that were not present with the other routines. 

The work of \citet{Jeffries2023} is a clear example of fully manufacturing a training dataset. This is the ultimate form of training data augmentation, and easily allows for a perfectly labelled training set, but comes at the cost that their network will inevitably perform differently (likely more poorly) on real-world data. It may be that the performance is still satisfactory, but extra care needs to be made to ensure that this is the case. Moreover, effort needs to be made to detect any odd behaviours with real-world data that do not occur with the artificial training data.

%

We next point out \cite{Kruk2022}. Like the work of \citet{Fuentes2010}, Kruk et al. also searched for streaks of minor bodies in HST data, specifically for data from the Advanced Camera for Surveys \citep[see][]{Ryon2023}, and the Wide Field Camera 3 \citep[see][]{Marinelli2024}. For training data, they made use of the Asteroid Hunter Project on the Zooniverse\footnote{\url{https://www.zooniverse.org/projects/sandorkruk/hubble-asteroid-hunter}} which presents HST imagery to a small army of volunteers who classify an image as containing a streak, and mark both the beginning and ending of a streak if present. Each image is classed by 10 different users resulting in high quality labels, though the sample size itself numbered only a few thousand samples. The training data and labels consisted of those images with asteroids, as well as masquerading bad sources which include cosmic rays satellite trails, and cosmic rays (appropriately labelled as such).

The authors make use of AutoML \citep{zoph2017} Vision Model.\footnote{\url{https://cloud.google.com/vision/automl/object-detection/docs}} This service aims to automatically explore different network architectures and select the most performant architecture for the problem presented. In some sense, AutoML aims to enable machine learning for non-experts, and is a good place to start for many budding ML users.

In their dataset \cite{Kruk2022} identify 1,300 asteroid trails, nearly half of which are from unidentified objects. Their network has a relatively low recall of 61\%. It is unclear if this low performance is a result of their training set (size or quality of labels), a failure of AutoML to find a highly performant network, or is simply due to the similarity in  morphology between asteroid trails and other non-asteroidal sources in the HST data. 

%

The DeepStreaks network is presented in \citet{Duev2019}. DeepStreaks is a custom network architecture designed to detect streaked fast moving minor bodies in ZTF imagery. DeepStreaks makes use of three separate classic networks as building blocks: ResNet50, VGG6, and DenseNet121 (see the Introductory chapter for a description of each network architecture ). A separate incarnation of building block network is trained to tackle binary classification of three separate but similar types of streak defined by their streak length. A logic branch exists for each problem class, which is the ensemble OR output of one of each of the different building block networks. The operation of the architecture is as follows. A given sample is passed through all nine separately trained building block networks. A candidate source is labelled as good if the candidate is labelled good by any one of the building block networks over two or more branches. This is an interesting idea as it presumably allows each network and each branch to be sensitive to its own particular style of streaked source, and so the ensemble should have better performance than its individual blocks or branches.

The training was performed in a supervised fashion, with train data assembled from a dataset made up of images with streaks previously identified by humans, synthetic streaks implanted into otherwise streak-less frames, and a set of imagery containing undesirable streak-like features such as cosmic rays and image ghost arcs. 

The outputs of DeepStreaks are still vetted by humans, but with a $50-100\times$ reduction in vetting of sources that would otherwise be false negative sources. A building block threshold of 0.5 probability for positive classification resulted in a $\sim97\%$ true positive rate.  The work of \citet{Duev2019} is an example of how classic networks which are simple to implement in modern ML frameworks can be manipulated in a straight forward fashion to produce an excellent outcome with easily understood operations. 

%

The work of \citet{Wang2022} is a direct competitor to the DeepStreaks work. They repurpose the EfficientNet-B0 architecture to search for streaks in two layer cutouts (image of the candidate streak and background template of the same region) of astronomical imagery. They generated a relatively hefty fully simulated training set, weighted 4:1 to those cutouts containing streaks and those that do not. They achieved an excellent low false positive rate of only 0.02\%. The EfficiencyNet-B0 model contains 4,161,268 trainable parameters. While \citet{Wang2022} make use of some techniques to avoid overfitting (see discussion in Section~\ref{sec:overfitting}), such as dropout, they did not present any effort to quantify overfitting. Given the complexity of the EfficientDet-B0 network, it seems likely that overfitting could be an issue. Whether or not this is more relevant to the performance of the network on real astronomical data compared to the fact that training data were fully simulated is unclear. 


We would like to point the reader to \citet{Lieu2019} and \citet{Varela2019}. \citet{Lieu2019} presents a good example of transfer learning, whereby they only retrain a subset of layers in a large and complex network that was previously trained on non-astronomical data. They consider a large range of available networks, and achieve a relatively good performance on asteroid streak detection even with a relatively small training set. The budding ML expert may start here, as this paper provides an excellent overview of the many things one needs to consider when embarking on a new ML project. \citet{Varela2019} present a good example of the application of the ``You Only Look Once'' or YOLO algorithm\footnote{YOLO was originally designed for fast facial recognition tasks} to the detect streaks in wide-field ground-based astronomy imagery. YOLO represents a very different approach to the use of CNNs compared to the other techniques discussed here.



The final paper I discuss is the work of \citet{Cowan2023}. This  tour de force in custom network design has aim to detect the streaks left by moving minor bodies. Specifically, they look for the blobby streaks left by bright asteroids in the mean stacks of spatially coincident imagery from the Microlensing Objects in Astrophysics (MOA) survey. The reason these streaks are blobby as compared to other efforts discussed above, is because MOA uses relatively short exposures in which asteroids remain nearly point-like, and so mean stacks images of asteroids appear as a nearly linear series of point-like sources.\footnote{They refer to these as asteroid tracklets. We call these streaks to avoid confusion with the astrometric tracklet terminology introduced above.}

\begin{figure}[h]
\includegraphics[width=0.95\textwidth]{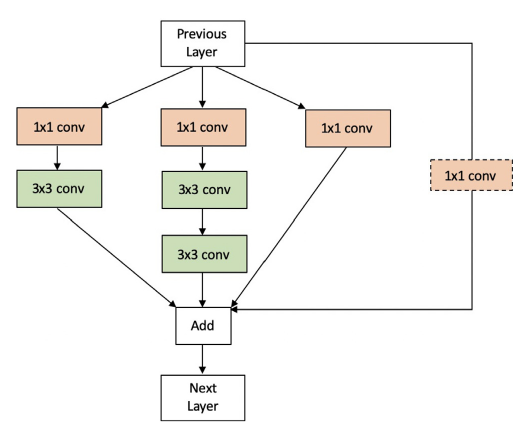}
\caption{The custom ResNet-style block invented and utilized by \citep{Cowan2023}. Note the difference with the standard ResNet block which sums only one convolutional chain with the skip connection. Reproduced from Figure~9 of that work. \label{fig:cowan-block}}
\end{figure}

\citet{Cowan2023} created 5 different custom networks that each outputs a likelihood that an input image contains a streak. Three of those networks are structured like a VGG with varying depths, and two are reminiscent of the ResNet, though these networks use a custom block which performs an addition of three separate convolutional layers of varying depth, and the ResNet skip connection (see Figure~\ref{fig:cowan-block}). While the performance of each network was tested individually, the best performance arose from the average ensemble of the outputs of all five networks. This is reminiscent of the results of \citet{Duev2019}, where the merger of outputs of a set of different network architectures resulted in good performance (though Duev did not discuss the performance of each individual network, only the logic-driven ensemble). The input to the network of Cowan is a set of 128x128 pixel cutouts selected from the full frame mean stacks, with output being a probability of the cutout containing a streak.

Training was done in a supervised fashion. Through a -- likely arduous -- manual search, a number of real asteroid streaks were identified and labelled. With augmentation (rotation, blurring, flipping, brightening and darkening), a large training set of cutouts was generated, weighted 1:5 with:without streaks. Critically, an independent test dataset was created from data of an entirely different night, and from some chips of the detector mosaic that were excluded from the training set. That is to say, their test set was entirely independent of their training set. This is a crucial feature of ML training that should be utilized when ever possible so as to preserve the veracity of the test results.

With this ensemble network, Cowan achieved 96\% recall and 87\% precision, which are extremely good results. It is unclear how the performance of this ensemble will degrade with asteroid brightness, an aspect not discussed in their paper. Though they make it clear from the start that their focus is the detection of bright (by some standards) asteroids with V<21.

%

\subsection{Detection of Moving Objects in Shift'n'Stack Imagery \label{sec:ML-sns}}
We now turn our attention to the application of CNNs to the good:bad binary classification of candidate sources detected by shift'n'stack type searches. The first application of this is from \citet{Smotherman2021} who apply a ResNet-50 network to vet the candidate detections produced by KBMOD (see Section~\ref{sec:shiftnstack}). The network accepted as input 21x21 pixel stacks centred on candidate sources. The stacks were produced during the shift'n'stack process. The network output is the probability of a candidate being real. Training of that network was done in a supervised fashion. They chose to create an artificial dataset of positive sources by injecting Gaussian point-sources along predetermined linear trajectories, with real negatives originating from outputs of actual search data. With this process they created a sample of 40,000 candidate sources, 70\% of which went to training. Details of the process are sparse, but they claim an accuracy of 96\% on the test set.

A weakness of the effort of \citet{Smotherman2021} is through the use of Gaussian point-spread functions (PSFs) for training. While these are somewhat point-like, they do not fully reflect the odd shapes that stars may take in astronomical imagery. As such, the performance on real astronomical imagery is inevitably reduced, though they do report a $\sim10\times$ reduction in the number of candidates per detector that require human vetting. We interpret this as $\sim10\times$ fewer false positives. A better choice would have been a Moffat profile \citep{Moffat1969} which more accurately reflects the shape of the core and wings of stellar PSFs. Better still would be to use an actual PSF model. 

I now turn my attention to the custom CNN I have developed to perform the same style of vetting as done by \citet{Smotherman2021}. That is, real:bogus classification of candidate sources found by shift'n'stack searches. 

The impetus for this effort came from the search for Kuiper Belt Objects bright enough to be observable by the LOng Range Reconaissance Imager (LORRI) on board the New Horizons Spacecraft \citep{Fraserlpsc2023}. The search field was at low galactic latitudes $\sim10^\circ$, with extremely high stellar density, that resulted in a greater than 1000:1 false positive rate in the KBMOD outputs of those search data, when searching to depths with sources reaching signal-to-noise$\sim5$ in a stack. This false positive rate saw little improvement with robust image subtraction. With thousands of candidates to vet per detector (the Hyper-SuprimeCam mosaic on the Subaru telescope has 103 available detectors), and $\sim 15$~epochs, human vetting was practically impossible, and so good:bad classification of the KBMOD stacks with CNNs was explored.

In my experimenting, I had first explored relatively shallow VGG-like CNNs. Generally, the performance matched the reported performance of the deep ResNet trained by \citet{Smotherman2021}, which reduced the human vetting workload by a factor of ten. Though to make the work practical, a further $10\times$ reduction was needed. Various custom modifications were explored, including networks that operated on only a finite range of source fluxes, networks that operated on a dual channel input of the mean and median KBMOD stacks, various versions of ensembles, and so on. 

\begin{figure}[h]
\includegraphics[width=0.95\textwidth]{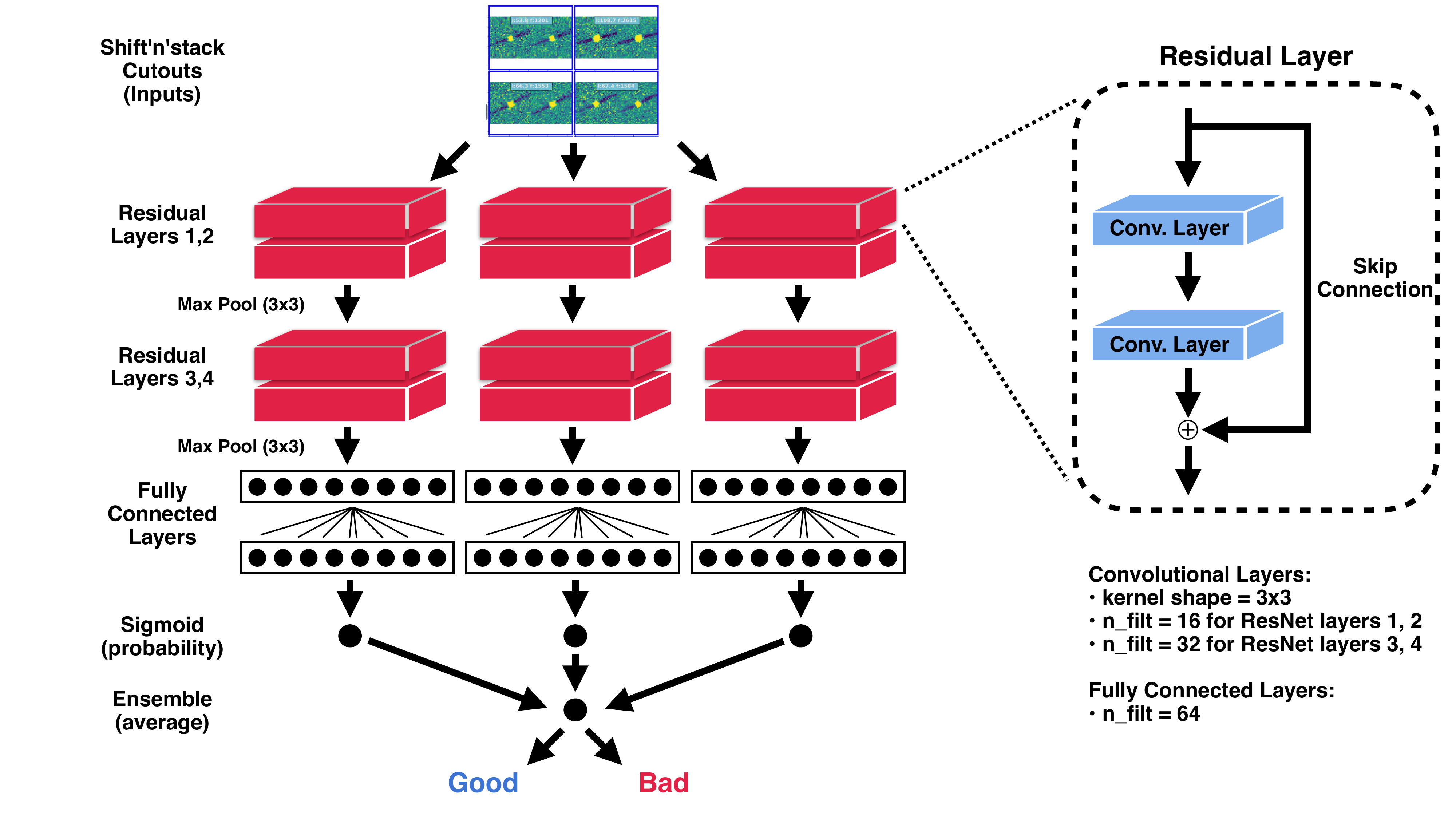}
\caption{The custom ResNet ensemble I have implemented for the binary classification of KBMOD shift-stacks. The filter parameters shown are those utilized in the New Horizons KBO search (see Section~\ref{sec:ML-sns}). Each of the three branches of the ensemble is trained on the same training set, but with a different random initialization of weights.  Reproduced from Figure~3 of an upcoming manuscript (Fraser et al. submitted to the Planetary Science Journal). We remind the reader to see Chapter~1 for an introduction to the ResNet and general CNN architectures. \label{fig:resnet}}
\end{figure}

The network that I settled on is an ensemble of ResNets. This network has proven extremely useful for the New Horizons Search, the CLASSY project, and in a deep search with the James-Webb Space Telescope that is being performed at time of writing this chapter \citep{Morgan2023,Eduardo2023}. A network diagram of one instance of the ResNet ensemble is presented in Figure~\ref{fig:resnet}. This network takes as input 43x43 pixel mean shift-stacks produced by KBMOD. 

Training is performed in a supervised fashion, and is done independently for each project (New Horizons search, CLASSY, JWST), as the detectors used in each project have vastly different image properties. Creation of a network that can handle data from different telescope facilities is an on-going effort. We present the results of this network applied to a small amount of CLASSY data.

A training dataset is randomly drawn from the KBMOD outputs. A key aspect to ensuring a high performing network was to train on samples that are morphologically as close to the real minor bodies as possible. To that end, imagery were populated by a large number -- roughly 100 per individual detector in a mosaic --  of artificial moving objects  on varied trajectories and over a range of brightnesses. A swarm of artificial Kuiper Belt Object orbits were generated, and propagated to the epochs of each individual image. These \emph{implanted} objects were injected into individual frames using a model PSF generated using the TRailed Imagery in Python package \citep{Fraser2016}, with the PSF modelled from point sources in the frames themselves. As a result, the injected artificial objects possessed morphology and noise properties that accurately reflected real minor bodies in the CLASSY imagery. KBMOD was executed on these implanted frames, from which a randomized training set was generated by labelling as good all implanted sources that were detected by KBMOD. All other KBMOD candidates were labelled as bad.  

Bogus sources were selected at random to build up a nearly balanced sample of 1.6:1 bogus:good sources. Effort was made to explore this ratio, as it was felt that increasing the bogus fraction would result in a broader variety of bogus sources to train on. It was found however, that any further imbalance beyond a factor of $\sim1.6$ in favor of the bogus sources, would result in a network that produced a massive fraction of false negatives, which needless to say would be catastrophic. This was true even with the use of sample weights to try to counteract the imbalance.

The above process of generating a training sample inevitably resulted in some real detected minor bodies being labelled as bogus. This mislabelling results in a $\sim0.02\%$ mislabelling of bogus sources. Such a small mislabelling of bogus labels did not influence the performance of the outputs. In the circumstance of a higher mislabelling, an iterative training procedure of relabelling between training may help. Rather I would recommend the training sample be augmented (e.g., with more implanted sources) to avoid the issue. The adage garbage in garbage out is true for most machine learning applications, and so the importance of maximizing the quality of the training set cannot be under-emphasized. 

By virtue of creating a training set from a KBMOD search of implanted imagery, the distribution of brightness of the sources labelled good inherently followed the distribution of implanted source brightnesses, convolved by the detection efficiency as a function of source brightness. In an attempt to avoid any bias this may introduce, we implemented sample weights for the good labelled sources that were the inverse frequency of occurrence of an object of that brightness in the detected sample. This was done by fitting a polynomial $f(m)$ to the histogram of brightnesses $m$ of implanted detected sources. The final sample weights of good sources was then $w(m) = \frac{1.6}{f(m)}$, and for bogus sources $w=\frac{1}{1.6}$.

Training of the network was otherwise commonplace, and made use of a cross-categorical loss, and the ADAM (version 2) optimizer, and 20\% dropout. The network itself was coded in keras. Augmentation of the training set involved linear shifts by one pixel in the four cardinal directions, rotations by 90, 180, and 270 degrees, and mirror flips vertically and horizontally. In my experience, these augmentations reduced the false positive rate of the network by $\sim10\%$. Surprisingly, I  found that augmentation by adding noise generally resulted in degradation of the network performance. I attributed this to the somewhat lazy approach I adopted, in adding Gaussian noise with standard deviation equal to that of the pixel values of the un-augmented sample. This created artificial samples with different noise properties than in the real data, and so forced the network to recognize a signal that was not present in real samples. A likely better approach would be to sample noise of the true background noise distribution, though I found that the training dataset was large enough to make use of this extra augmentation method unnecessary.

In practice, an instance of the network consisting of three branches was trained on a single training set, with each  branch trained separately using a different random weight initialization. In my experimenting, I discovered that the quality of network output in a training run could be highly variable, a problem that I was unable to solve directly. When a training run succeeded -- it did not always succeed -- I found that a single branch of the ensemble would better reject certain types of bogus source better than other branches. Generally, the average ensemble outperformed the individual branches that made up the ensemble. I found that three branches was the sweet spot in terms of keeping the number of trainable parameters low while ensuring a well-performing network, with diminishing returns for higher numbers of branches.

Due to the roughly 100:1 ratio of bogus to good sources that are output by KBMOD (when searched to a low SNR) a single randomly drawn training set never fully covered the broad range of morphologies exhibited by the KBMOD outputs, at least for the New Horizons KBO search. In that case, we found that an ensemble of three independent instances produced a satisfactory result. Each of three instances of the network shown in Figure~\ref{fig:resnet} was trained on a different randomly generated training set. We trained roughly 10 different unique instances of that network, and selected the three best performers for production. The final classification was done using the average ensemble probabilities of those three instances (each of which itself is an ensemble). 

\begin{figure}[h]
\includegraphics[width=0.955\textwidth]{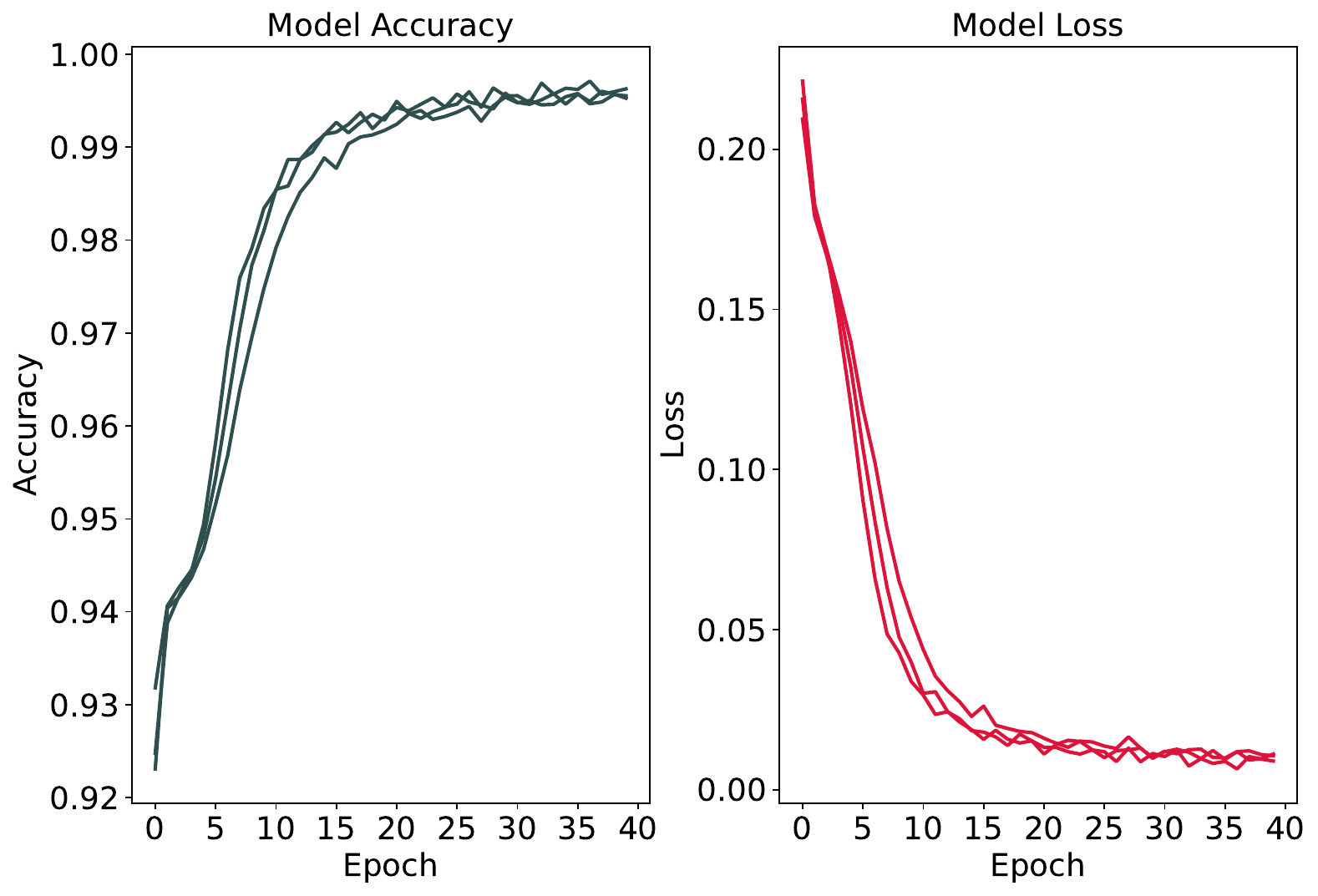}
\caption{Example of loss and training evolution during the training of one instance of the network depicted in Figure~\ref{fig:resnet}. For this run, training data were selected from 6 nights of data from the CLASSY survey, resulting in an augmented training sample size of 317,205. \label{fig:resloss}}
\end{figure}

\begin{figure}[h]
\includegraphics[width=0.95\textwidth]{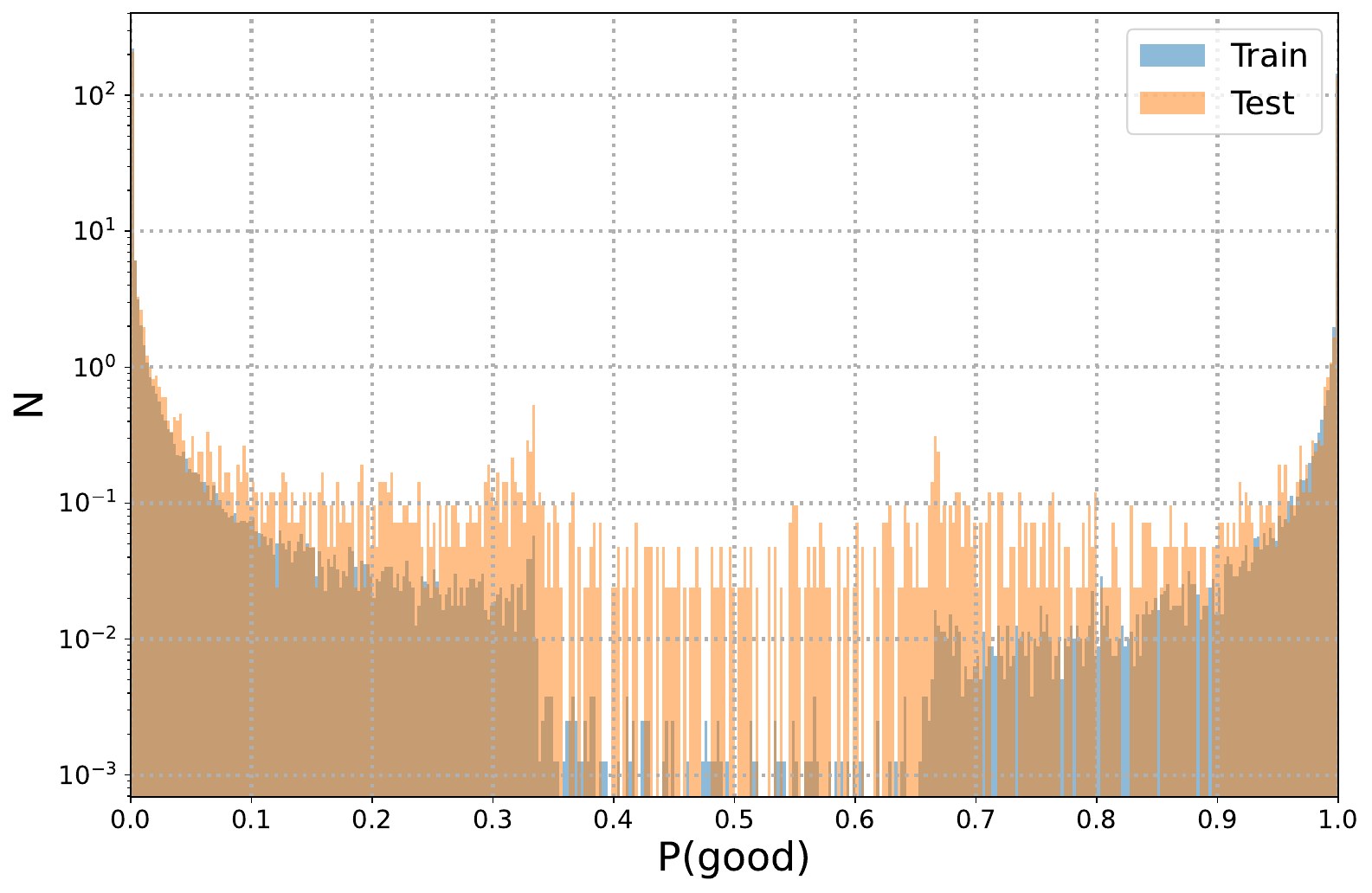}
\caption{Example of the prediction probability distribution of the training and test distributions of the training run shown in Figure~\ref{fig:resloss}. For readability, the training distribution is normalized so that the peaks of the training and test distributions match. The test sample (orange) shows a slightly higher fraction of samples with middling probabilities $\sim40$ to $60\%$ that is not present in the training sample. The general similarity of the training and sample probability distributions however, implies a successful training run resulting in a similar behaviour of network outputs for both samples. \label{fig:resprobs}}
\end{figure}

\begin{figure}[h]
\includegraphics[width=0.95\textwidth]{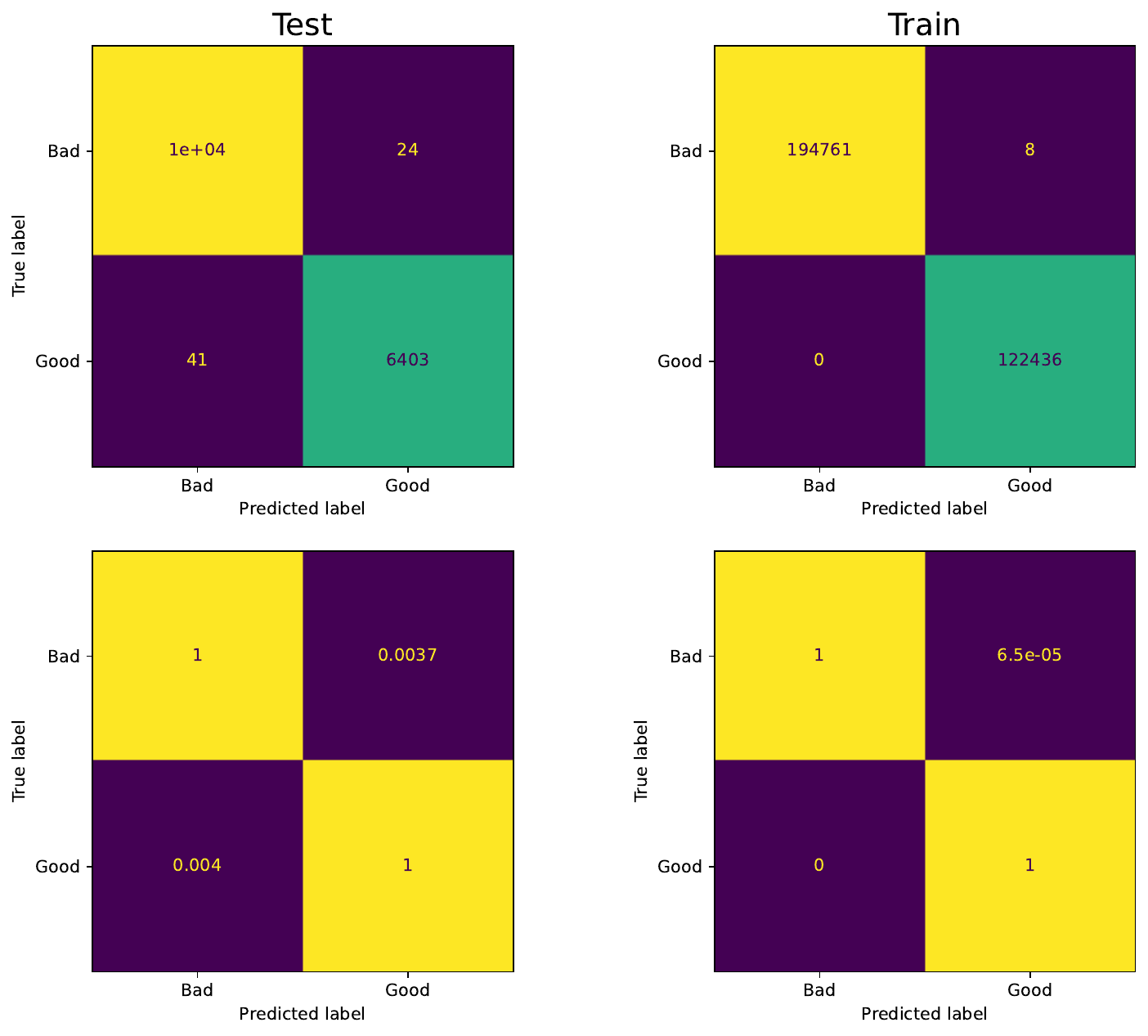}
\caption{Quadrant plots for the training and test samples used in the training run shown in  Figure~\ref{fig:resloss}. Absolute and fractional versions are shown. The test sample shows a higher fraction of false positives than occurs for the train sample, implying a small amount of overfitting has occurred.  \label{fig:resquadrant}}
\end{figure}

\begin{figure}[h]
\includegraphics[width=0.95\textwidth]{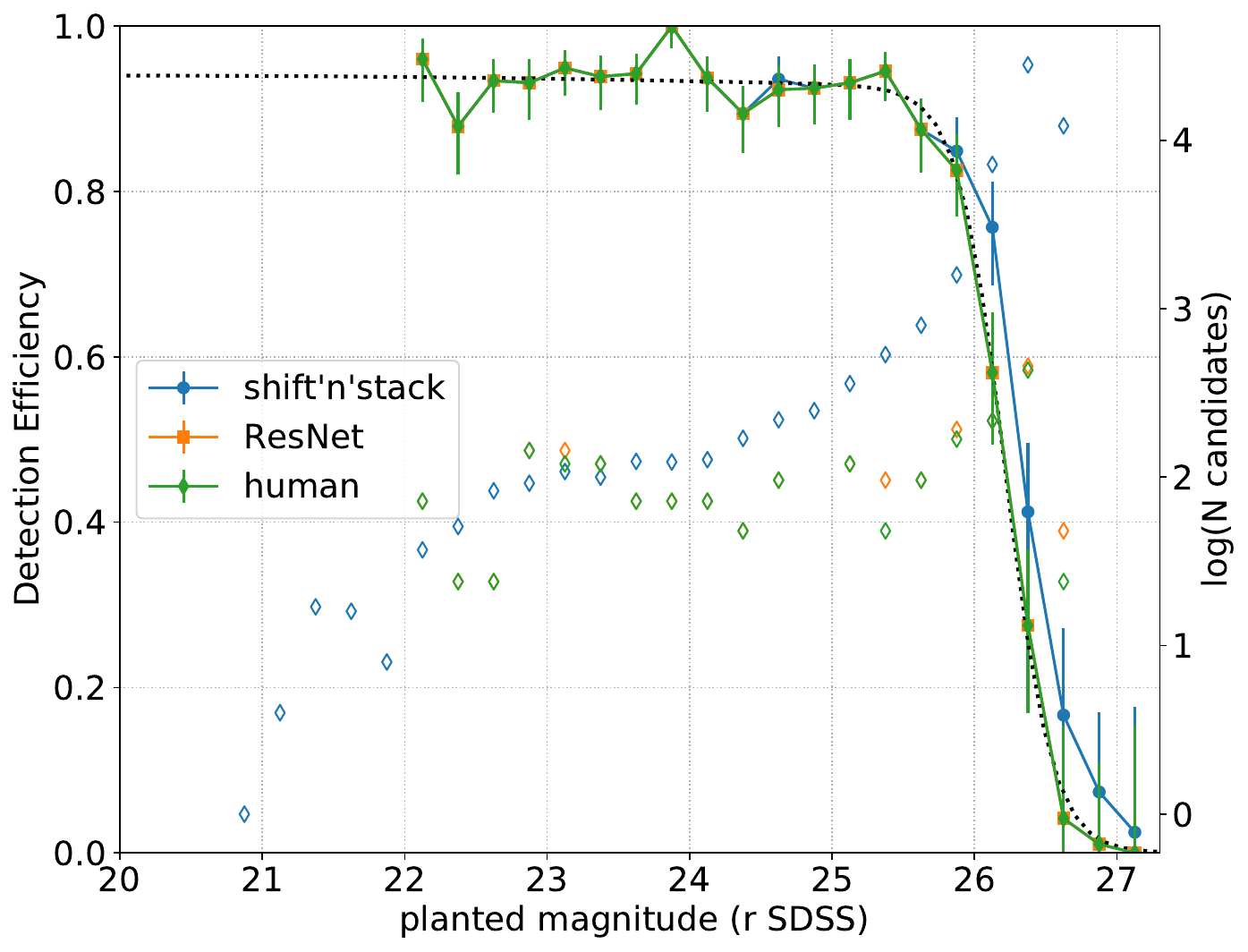}
\caption{Example detection efficiencies of the KBMOD outputs (blue), those labelled ``good'' by the ResNet ensemble (orange), and those finally labelled as ``good'' by a human (green). These are evaluated against the artificial implanted moving sources. Errorbars are 1-$\sigma$ poissonian intervals. The black dotted curve is a efficiency curve fit of the form $\eta(r) = \frac{0.94-4.1\times10^{-4} (r-20.0)^2}{1.0+\exp((r-26.2)/0.16)}$ to the human vet outputs. The open diamonds plot the number of candidates output at each stage of the vetting, with colours matching the efficiency curve values. Note - the lack of efficiency curve data brightward of r=22 is an artifact of not implanting artificial objects brighter than this value. \label{fig:classy-eff}}
\end{figure}

In Figures~\ref{fig:resloss}, \ref{fig:resprobs}, \ref{fig:resquadrant}, and \ref{fig:classy-eff}, I show examples of the diagnostic plots I made most use of when determining the performance of the network. Figure~\ref{fig:resloss} shows the behaviour expected of a successful training run: respectively, rapid increase and decrease in model accuracy and loss over the first few epochs that stabilizes on low values of loss and accuracy approaching 100\%.

Figure~\ref{fig:resprobs} shows the prediction probability distribution of the training and test datasets. I have found this particularly useful in checking for a failed training, and for significant overfitting. In the case of a failed training, the distribution will not cleanly separate into two classes at some P value as it does here. Moreover, a well performing network will result in a wide and nearly empty valley between those samples that cluster near P-values of 0 and 1. In the example shown, there is a hint of overfitting evidenced by peaks of samples with middling probabilities at $P\sim0.3$ and $P\sim0.7$ that are more prominent in the test data than they are in the train sample.

Figure~\ref{fig:resquadrant} presents the usual quadrant plots for binary classification for both train and test datasets. The performance is generally good, with precision of 98.7\% for the test sample, and 99.9\% for the train sample and recall of 99.4\% and 99.9\% for the test and train samples, respectively. As is common with ML efforts, the test dataset does not perform quite as well as the training dataset, in this case only marginally. Though I point out that a better test of network performance will come from the application of the network to a dataset from a different night than used for training and testing.  This is presented in Figure~\ref{fig:classy-eff}, which shows an example of application of the above network ensemble applied to data not used to assemble the training or test datasets. The training and test datasets included data acquired on nights of August 22, 23, and 26, 2022. Data from August 24, 2022 were acquired in similar transparency and image quality as neighbouring nights, and are an ideal validation dataset.

Figure~\ref{fig:classy-eff} shows the typical behaviour experienced in the CLASSY search. The KBMOD outputs have the faintest search depths, but report back $94,722$ candidate sources across the 40 detectors of MegaCam, $2,088$ of which are planted and $\sim70$ are real minor bodies. Application of the ensemble reduces the search depth by $\sim0.1-0.2$ magnitudes, and the sample labelled good is a factor of $\sim25\times$ smaller than the KBMOD detection list, at $3,628$ sources. Final human vetting results in a further degredation of the search efficiency, the magnitude of which depends on the performance of the network on the particular night in question, and can sometimes result in $\sim0.2$~mags loss in search depth. In this example, of the 3,359 sources labelled good by both human and network, 2016 and planetd, and $\sim70$ are real minor bodies. 
The remaining $\sim1,200$ sources are false positives, with the vast majority having SNR$\sim5$, or detection efficiencies $\eta<0.4$ where both machine and human can no longer distinguish real sources from noise peaks. The remaining false positives are ultimately culled at the tracklet creation stage where three nights of data are grouped together (see Section~\ref{sec:shiftnstack}). At this point the detection list purity is essentially 100\% \citep{Fraser2023acm}. The New Horizons search also relied on linking to cull the remaining false positives, though the search had to be limited to a higher SNR due to the significantly higher prevalance of subtraction residuals which resulted in significantly noisier stacks. 

\section{Source Brightness Regression \label{sec:mag-regression}}
In this section I discuss the use of a CNN\footnote{see Chapter~1 for an introduction to the CNN} to quickly estimate source brightnesses from the shift'n'stack outputs discussed in the previous section. The network I discuss below can produce photometry estimates orders of magnitude more quickly than PSF fitting techniques with similar precision. The main strength of regression via CNN however, is that the network can be trained to learn what sources of flux in a stack correspond to the minor body, and what other sources in the stack are superfluous (noise, background, etc.), as long as the sources in the training set contains said superfluous sources. That is to say, through basic training, in most circumstances the CNN will automatically handle effects of contamination in the stacks that would be extremely difficult to manage through classical PSF fitting methods. 

We make use of the methods presented in \citet{Bialek2020} whereby with clever use of an appropriate likelihood function, the network can be trained to predict a brightness value as well as an estimate of the brightness uncertainty, \emph{even when photometric uncertainty is not part of the information available to training}. The network is trained to estimate the probability density function (PDF) of a source's flux from that source's input, and the mean and standard deviation of that probability distribution are returned. As we are measuring fluxes of background limited sources, we are justified in adopting the negative log-likelihood for a Gaussian distribution as the functional representation of that probability density function, which is given by

\begin{equation}
-\log p_\theta\left(y|x\right) = \frac{\log \sigma_\theta^2\left(\bm{x}\right)}{2} + \frac{\left(y-\mu_\theta(\bm{x})\right)^2}{2\sigma_\theta^2(\bm{x})}.
\end{equation}
\noindent Here, $y$ is the flux value of stack $\bm{x}$. The $\mu_\theta(\bm{x})$ and $\sigma_\theta$ are the network estimates of the mean and standard deviation of the flux PDF, $p_\theta\left(y|x\right)$. $\theta$ are the learned weights of the network.

To accommodate training variances, an ensemble of $M$ predictor networks are trained independently. The final PDF is simply the average of the PDFs of each network

\begin{equation}
    p\left(y|\bm{x}\right) = \frac{1}{M} \sum_{m=1}^{M} p_{\theta_m}\left(y|x\right)
\end{equation}

\noindent and so the ensemble estimate of the flux is just the mean of flux estimates of each of the networks

\begin{equation}
    \mu_*(\bm{x}) = \frac{1}{M} \sum_{m=1}^{M} \mu_{\theta_m}(\bm{x}).
\end{equation}  

\noindent The final variance on the flux estimate is given by 

\begin{equation}
    \sigma_*^2 = \frac{1}{M} \sum_{m=1}^{M} \left( \sigma_{\theta_m}^2\left(\bm{x}\right) + \mu_{\theta_m}^2     \right) - \mu_*^2(\bm{x}).
\end{equation}

\noindent The above includes errors induced by training variance (aleatoric uncertainty), and variance due to the imagery themselves (epistemic uncertainty). 

For this example, we use a shallow, run of the mill VGG-like CNN, with only three convolutional layers that feed two percepteron layers. A CNN as the starting point for this effort seemed obvious given the image nature of the data inputs. I have found that such a simple network is sufficient for the task at hand. As a result, no further effort was expended to search for a more complex architecture such as a ResNet, though there are a few remaining issues we discuss below that may be solved with a more complex architecture.

In the network I settled on, the first two convolutional layers have only 8 filters with the third only 4, and the percepteron only 16 neurons.  In all cases, max pooling is used after the first CNN layer. The activation on $\sigma$ is a modified \emph{elu} of the form 

\begin{align}
f(x)&=\exp(x)+10^{-16} \mbox{ for $x<0$} \\\nonumber
    &=x \mbox{ for $x\geq 0$}.    
\end{align}
\noindent
Inputs to the network were the same 43x43 pixel stacks used in the ResNet-based binary classifier I discuss above, and the known \emph{implanted} flux values for each stack. For this example, we adopt $M=5$.

In developing this example I made use of the advice I give below in avoiding overfitting. In particular, I first experimented to find a CNN that produced satisfactory results, and then experimented with its complexity to find a network that had a low number of trainable parameters while maintaining satisfactory outputs, and to search for the effects of overfitting. In particular, I explored two additional networks with the same architecture as above, but with double and with half the number of convolutional filters and neurons.

The implanted sources in four nights of CLASSY data from August 1, 22, 23, and 26, 2022 were utilized to make the training set. Implanted sources detected with KBMOD were used, with their known instrumental fluxes -- apparent fluxes were inappropriate because night to night variations in the flux zeropoints would cause massive variations in the mean predictions. The training set was augmented with 1-pixel linear shifts in the four cardinal directions, mirror flips in the horizontal and vertical axes, and rotations at $90^\circ$, $180^\circ$, and $270^\circ$. After augmentation, the training sample was 95\% of available sources, or 260,544 samples, and the remaining 13,696 used for test.

Training was done using the ADAM optimizer, batch size of 4096, and sample weights as the inverse of the source flux distribution were used, like with the ResNet binary classifier, though we point out that for this case, regressor performance was not appreciably altered with sample weights. 

\begin{figure}[ht]
\includegraphics[width=0.9\textwidth]{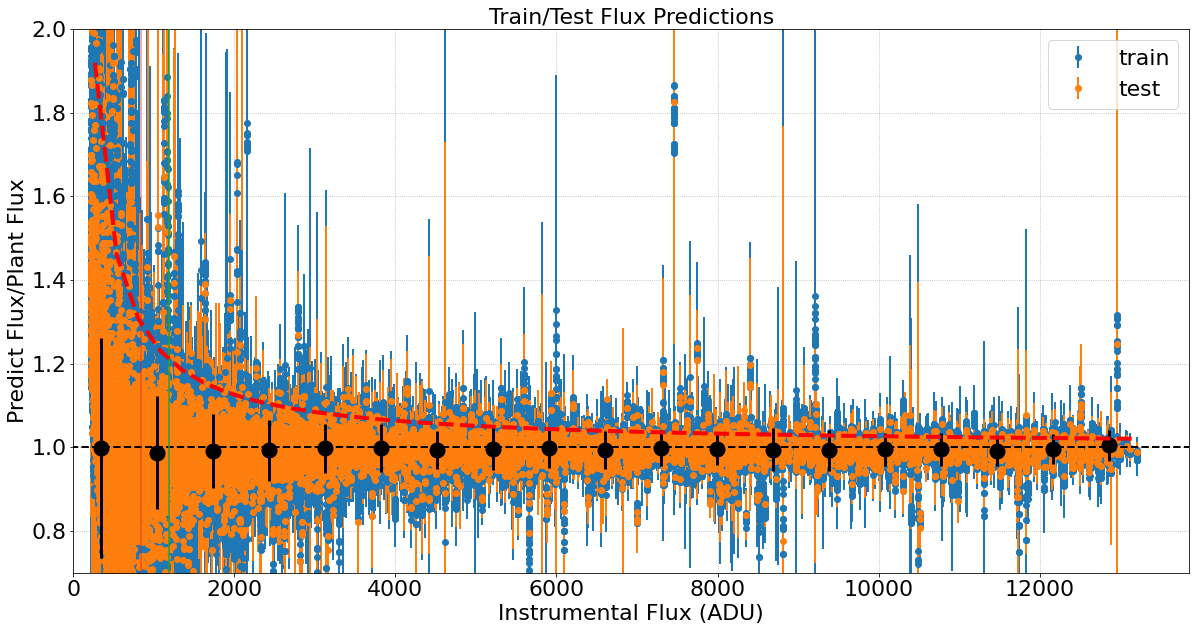}
\includegraphics[width=0.9\textwidth]{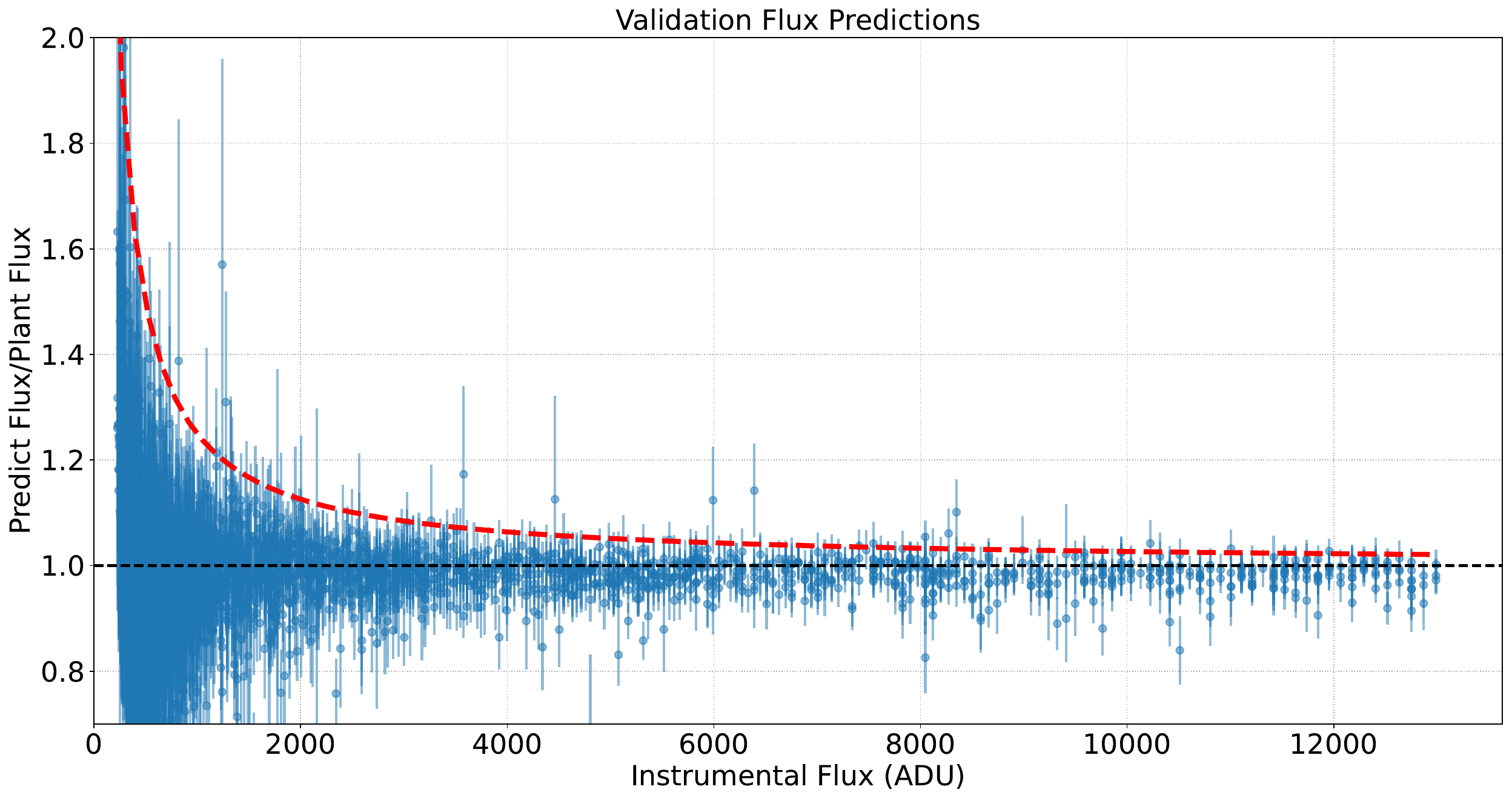}
\caption{\textbf{Top:} The ratio of predicted to true instrumental flux of implanted sources as a function of planted source instrumental flux for the train (blue) and test (orange) dataset -- the test sample was 5\% of the available training sample. Black circles show the mean predicated/planted flux ratios of the test set in 20 bins spanning the range of implanted fluxes. The errorbars on the means show the mean predicted uncertainty on the instrumental flux of the samples in each bin. The red curve shows the SNR expectation for background limited aperture photometry. \textbf{Bottom:} The ratio of predicted to true instrumental flux of implanted sources in a validation dataset which is generated from an entirely different epoch of data than the four used to generate the training and test datasets. The red curve is the same as shown in the top panel. \label{fig:fluxregressor}}
\end{figure}

In Figure~\ref{fig:fluxregressor} we present a ratio of the predicted instrumental flux to true instrumental fluxes for implanted sources. We also show an estimate of the SNR that would be achieved with aperture photometry. For that we consider a circular aperture with radius $1.2\times$ the typical Full-Width at Half Maximum for CLASSY imagery, or $\sim0.6"$ which corresponds to  an aperture area of 68 pixels, and we adopt a background value of 900 ADU. As can be seen, the predictions broadly match the precision one would achieve with well tuned aperture photometry \citep{Fraser2016}. Furthermore, the output exhibits no appreciable bias down to the level of flux where Malmquist bias clearly sets in, showing less than 0.5\% bias for test, training, and validation datasets, for all three networks.

\begin{figure}[ht]
\includegraphics[width=0.9\textwidth]{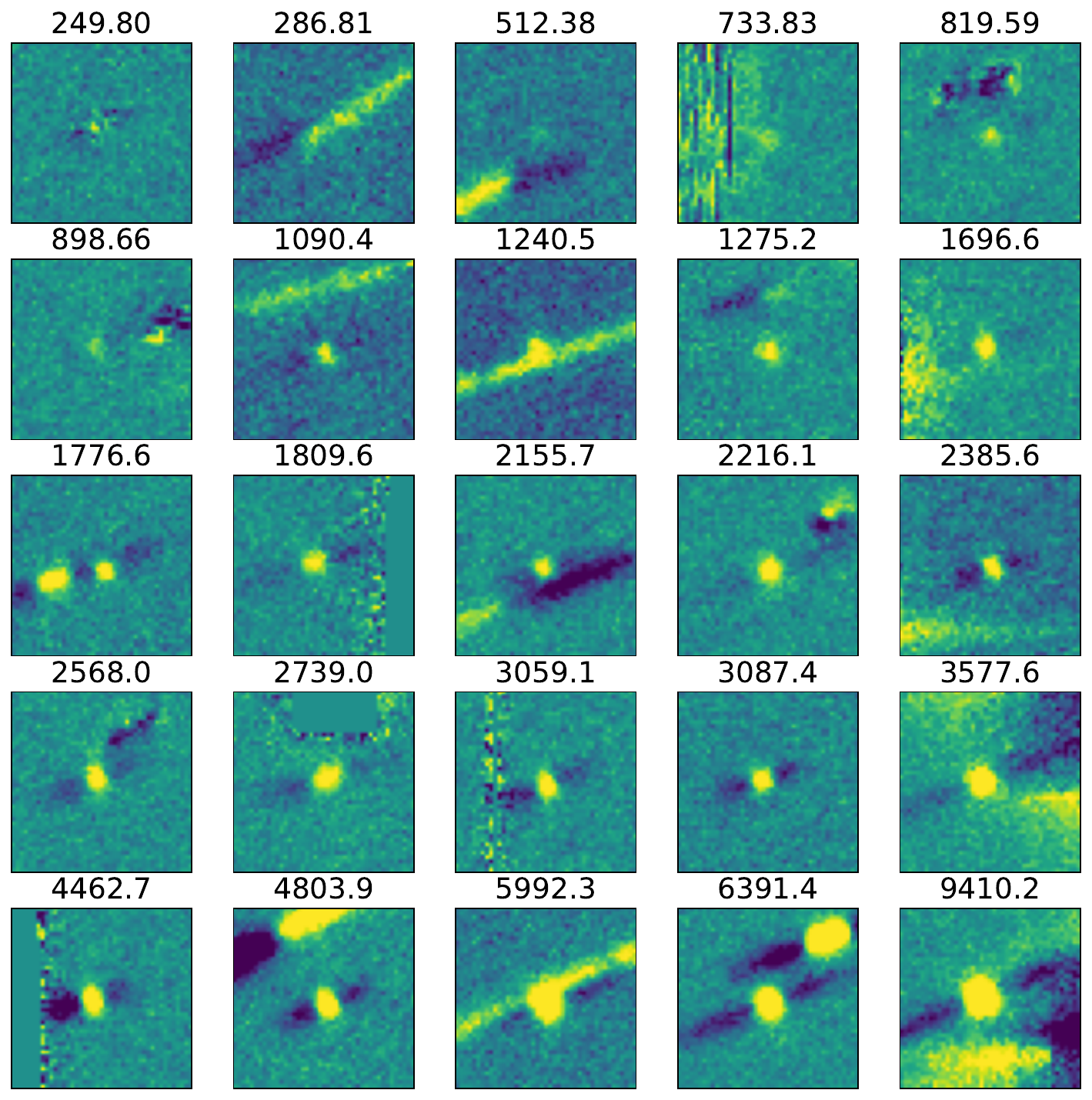}
\caption{ The 25 sources with the most discrepant flux predictions, which represent $\sim1\%$ of the validation dataset. Background contamination is an issue for the regressor model. The predicted flux for each source is shown above the stack. Most of these sources are relatively bright, though contamination is similarly prevalent for fainter sources, but uncertainties of those fainter sources are appropriately larger, and so the effects of contamination are relatively less influential. \label{fig:outliers}}
\end{figure}

For validation, we utilized the KBMOD outputs for a night of CLASSY data not utilized in creation of the training or test datasets (see bottom panel of Figure~\ref{fig:fluxregressor}). The validation data are the same as those used to generate Figure~\ref{fig:classy-eff}, and show the same performance as for the test and training data, with flux predictions scaling appropriately with expectations from SNR estimates. Though the validation data show lower uncertainties in the flux predictions, measured both by the scatter in true minus predicted fluxes as well as the predicted uncertainties, $\sigma_*$. This is likely due to better observing conditions experienced during the observations used for validation compared to those used for training. 

From the black points in Figure~\ref{fig:fluxregressor} the predicted uncertainty $\sigma_*$ is typical of the true scatter of the predicted fluxes around their true values.  Moreover, the distribution of scatter in predicted fluxes roughly follow expectations of a normal distribution; 6 of the 2,295 sources have predicted fluxes more than $3\sigma_*$ from their true values, where we would expect 7.

It can also be seen that for a very small fraction of samples, the predicted uncertainty is much higher, and the predicted flux is incorrectly predicted.  These \emph{outliers} were identified by comparing each source's predicted, $\sigma_*$ to those values of the 50 neighbouring sources when sorted by implanted flux (not predicted flux). I found that twenty five sources had $\sigma_*$ twice the median value of the 50 neighbours. These are shown in Figure~\ref{fig:outliers}. These outliers are flux measurements of those shift-stacks that are contaminated by other astronomical objects in the frame, or in the case of lower than correct predictions, contamination by negative subtraction residuals of nearby bright stars. It is clear however, that those sources with the most incorrect predictions are also those with typically the largest fractional uncertainties.

In a search for overfitting, I compared the outputs of the nominal model to that of the large and small models, which have 1,896, 7,098, and 896 trainable parameters for each of branch of the ensemble, respectively. For all three models, the number of outliers which were chosen as $\sigma_*$ twice the median value of the 50 neighbours, was comparable between models, with 29, 25, and 36 outliers for the small, nominal, and large models, respectively. The small model identified the same outliers as the nominal model, but also included four of the brightest sources which all had underestimated flux predictions. The brightest sources all have significantly different morphologies compared to the fainter sources, with much deeper and longer negative subtraction wells compared to the significantly more numerous fainter sources. It seems the small model possesses too few parameters to maintain the capability of handling the full range of morphologies in the datasets. The large model similarly reports many bright sources as outliers in addition to the same 25 outliers reported by the nominal model. In this case however, the large model reports appreciably smaller uncertainties for bright sources: for sources with flux $>6,000$~adu, the median $\sigma_*$ is 0.05, 0.04, and 0.03~mags for the small, nominal, and large models, respectively. From a noise perspective, $\sim0.04$~mag uncertainty is expected for this brightness range, and so it seems the largest model is overfitting for the brightest sources which are relatively rare in the training datasets, resulting in an unreasonably small $\sigma_*$ for those rare sources.

The network I used for this flux regressor is extremely simple compared to typical CNNs seen in the astronomy literature. It is shallow, with very few filters, and as a result, possesses a very small number of trainable parameters. But with an excellent training sample, and appropriate selection of PDF, the performance of the network is admirable. 

The result of being able to create a network that can predict both flux and its uncertainty may seem paradoxical given that no information about flux uncertainty was provide at training. The trick is with the use of a likelihood that is appropriate for the measurement in question -- in this case a Gaussian for flux -- that itself also depends on a parameter that encodes variance -- in this case $\sigma$ of the Gaussian.  One can create a network that provides an estimate of the uncertainty on the regressed parameter, seemingly for free. 

In summary, it seems the nominal model produces uncertainties that match expectations of photon counting, as measured by values of predicted $\sigma_*$, as well as the number of $3\sigma$ outliers in the validation dataset (once contaminated outliers are removed). This analysis demonstrates that overfitting is not a major issue for the nominal network, as might be expected; the training sample size of $\sim260,000$ samples (after augmentation) is more than two orders of magnitude larger than the 1,896 trainable parameters in this simple network. The larger model shows signs of overfitting resulting in overpredicted precision for bright sources. The comparison of the small, nominal, and large networks demonstrates the effort typically required to reveal any overfitting, which we discuss further in the next section.

This regressor is not without its weaknesses. The main outstanding issue is with management of those sources with incorrect flux predictions as a result of background contamination, which only represent $\sim1\%$ of this validation sample. I have yet to solve this issue. I believe the underlying cause is likely a result of the contamination which largely invalidates the assumption of Gaussin PDFs for these sources. This suggests a possible recourse whereby these outlier sources are first identified by their unreasonably large uncertainties compared to other similarly bright sources, and their fluxes remeasured with an additional network trained with a PDF likelihood that is more appropriate for contaminated imagery. Future experiments are required to determine the best form of this likelihood.  

The three networks discussed above, as well as associated data-files, and an example notebook showing the training and usage of those networks is available from the GitHub repository associated with this textbook (see Section~\ref{sec:code}).

\section{Overfitting \label{sec:overfitting}}
Here we talk about overfitting, and its avoidance. Much of the advice I give in this section comes from practical experience and not necessarily from literature examples, though we point the reader to already discussed papers \citet{Cowan2023} and \citet{Duev2019} for some discussions on overfitting avoidance. Overfitting remains an active area of research in machine learning.

Overfitting is the circumstance in which a network seemingly returns outputs or demonstrates performance on training data that differs significantly to outputs from data not involved in the training. In my opinion, this is one of the most nefarious and damaging potential consequences of the use of machine learning for scientific purposes, and is unfortunately not often considered -- not all of the papers I summarize in Sections~\ref{sec:ML-applications} even mention the word ``overfitting''!

Consider this particularly scary, and unfortunately true anecdote. A network was being trained to detect moving objects in multi-frame inputs of astronomical imagery, whereby any moving objects are not at the same position from frame to frame. Using implanted sources like those that I have discussed in the previous section, the network was trained to great success. During experimenting, a bug was introduced resulting in only the first frame being processed by the network, and yet after training, a magical result occurred in that the network was still able to find the moving sources in just those singular frames. With effort, it was realized that the network was trained too deeply and instead of learning to detect moving sources, it instead learned to detect subtle features unique to the artificial implanted sources that were not found in real point sources, and was not finding real moving sources at all. Such a network behaviour would have been catastrophic if gone unnoticed and the network were put into production. 

The above anecdote provides a good lesson for ML users both novice and expert and is a stark demonstration of the dangers of overfitting. It is easy to imagine a less fantastical result than being able to detect moving sources in single frames that would slip past the attention of even an experienced ML user.

Overfitting itself is simply enabled by the huge complexity of many modern neural networks. Consider for example, VGG-15, with 138 million trainable parameters. Such a complex network needs a sufficiently complex training set, or otherwise during training the network could learn to identify individual samples in the training data rather than meaningful features for the problem at hand (e.g., morphology) that distinguish different classes of source.

Guidance on when a circumstance might allow for overfitting can come from similar lessons learned in classical chi-squared curve fitting. The reduced chi-square considers the degrees of freedom $\nu$  in the problem,  where $\nu = N_{samp} - N_{param}$, and $N_{samp}$ and $N_{param}$ are the number of samples used in the fit, and the number of free parameters in the model, respectively. For a meaningful curve fit, $\nu$ should be very large. In a similar way we can approach neural network training. That is, ensure that the number of samples in a training set is significantly more numerous than the number of trainable parameters. Using $\nu$ as a metric, the ResNet binary classifier I discuss above had $N_{param}=9.5\times10^5$ trainable parameters (316,354 per branch of the ensemble), while the number of samples (after augmentation) used in training was 317,205. Each sample is itself a 43x43 pixel image, and so the total $N_{samp}=0.56\times10^9$, or $\sim600\times$ larger than $N_{param}$. This is a large difference which suggests that overfitting in this circumstance should not be an issue, though as I point out above, there are some hints of slight overfitting that has occurred, and so this comparison of $N_{samp}$ to $N_{param}$ should only be used as guidance. In the flux regressor example, the $N_{samp}=0.5\times10^9$ while $N_{param}=35,490$. With such a huge training sample compared to the trainable parameters, it is unsurprisng that I have not been able to find evidence of overfitting for the regressor.

Evidence for overfitting in the ResNet is seen in Figures~\ref{fig:resprobs} and in \ref{fig:resquadrant} with each showing slightly worse behaviour for the test dataset than the training dataset. In some circumstances the validation dataset may be useful for identifying overfitting, most likely in the form of worse network performance on the validation data.

Direct tests for overfitting are difficult and circumstantial. For the binary classification problem, a robust test could be to create two \emph{fully} independent training sets, train a network on each training set, and then cross validate the outputs of the training sets on the networks they were not used to train. To ensure fully independent training sets one would need to utilize two independent observational datasets, with independently implanted artificial sources, ideally using different PSF models (e.g., TRIPPy PSFs and noiseless Gaussian sources), and which used independent templates for background subtraction. Then one could be confident that the networks trained on each train set would also be independent \citep[see][for an example of this approach]{Pontinen2023}. Clearly the burden for such a check is very high. An equivalent procedure would need to be devised for the flux regression problem, as it would for any different ML application. As of yet I have not seen a robust general method for checking of overfitting in CNN training.

There are many techniques to avoid overtraining. The first and most obvious is to ensure one is using a massive and high quality training set. The creation of a high quality training set is often a large undertaking, and should not be avoided. Steps should be taken to maximize the sample size. Augmentation can greatly help, and there are many techniques that can be impleneted, including translations, rotations, mirroring, noise addition, contrast changes, etc., though with the addition of each new augmentation one has to check that network performance is not degraded -- a common occurrence with the addition of noise, for example. If the training set is to be artificially generated, effort must be made to ensure that the artificial positives match reality as close as is possible, or one should expect significantly degraded performance in production settings. We direct the reader to \citet{Hausen2020} for a short discussion on overfitting avoidance through augmentation.

Effort can be made during training to prevent overfitting. For example, one can halt training before training converges on a minimum loss value (using the keras early stopper callback, for example). This can help prevent the training from reaching a local minimum that can produce unreasonably confident classification probabilities for the training data. Tuning the stop time is challenging however, because too early and the network performance can be degraded. Tuning early stopping remains very much an art. Furthermore, one should use dropout, which randomly selects some fraction (usually 15-25\%) of trainable parameters that are not adjusted at a given training epoch. This approach can greatly help in avoiding local minima in the loss phase space which are a main cause of overfitting. Batch normalization can also help in overfitting avoidance.

If a pretrained network is used as a starting place, then one can train only certain sections (e.g., the last fully connected layer of a VGG network). This can keep the number of trainable parameters low, and reducing the chances of overfitting. 

Finally, overfitting can be reduced at the network design level. One is almost always better off to use a less complex network (fewer layers, fewer filters, etc.) than a more complex one, as long as the networks show equal performance. Once one has found a general network architecture that works for an application, an iterative process of reduction in complexity and retraining should be performed to find the minimal network that still retains satisfactory performance. This is the principle technique employed by myself and by \citet{ChybaRabeendran2021} to avoid overfitting.

\section{Applications of Machine Learning In The Era of Big Data Surveys \label{sec:future}}

Here I conclude with some speculative remarks about the applications of machine learning in future large scale astronomy projects. The future prospects for the observational science of minor bodies is exciting, with many large scale surveys on the near horizon, including Euclid, and the Vera C. Rubin Legacy Survey of Space and Time.

After a successful journey to the L2 Lagrange point, Euclid has begun its visible and near-Infrared (NIR) survey \citep{Laureijs2011} which will cover nearly 15,000 square degrees of the sky in the I, Y, J, and H filters, as well as provide slitless spectroscopy of much of that region. It is expected that Euclid will observe roughly 150,000 minor bodies \citep{Carry2018} to a depth unheard of in NIR surveys -- Euclid should achieve a limiting magnitude of $m_{AB}\sim24.8$ in all three NIR filters . Study of how best to detect and extract minor bodies in Euclid imagery is already on-going. Using simulated Euclid observations, \citet{Pontinen2023} studied the utility of CNNs to identify streaked images of minor bodies in Euclid data. This fairly advanced technique uses CNNs to identify streaks in small cutouts, a recurrent neural network to stitch together streaks spanning neighbouring cutouts or images, and then a gradient boost algorithm to identify start and end points of the merged streaks. Their procedure outperforms the non-ML version of a competing algorithm in virtually every way: a $2.6\times$ reduction in run time; doubling of purity of the outputs; and a $\sim0.5$ magnitude improvement in brightness of detected streaks. This impressive work presents the sort of efforts that are required to effectively search massive datasets like that currently being produced by Euclid. Also see the work of \citet{Lieu2019} which we discussed earlier.

The Euclid observing strategy is essentially is to acquire four 600~s dithered exposures in 4,400 seconds \citep{EuclidCollaboration2022}. Due to spacecraft pointing constraints, observation geometry is always essentially near quadrature, which naturally provides a circumstance of low apparent velocities for minor bodies in most of the Euclid pointings. For KBOs for example, the rates of motion should be no more than a few tenths of an arcsecond per hour, resulting in total streaks of no more than a few pixels in length across the entire 4,400~s sequence. At time of writing, I am unaware of any publications discussing the detection of slow moving -- untrailed --  minor bodies in Euclid imagery. 

It is likely that Euclid could benefit greatly from a technique like that presented by \citet{ChybaRabeendran2021} to look for KBOs and other slow moving minor bodies. One significant improvement that seems necessary for efficient search of Euclid data is to forgo the use of cutouts, and rather make use of larger image regions, and to make use of a CNN that can manage the actual detection of moving sources within the frames. This would skip the expensive computational step of first producing a list of candidates as is currently done with ATLAS data, and may afford an even deeper search than more classical non-ML methods could provide.

With first light expected in mid-2025, over the following ten years, the Legacy Survey of Space and Time (LSST) will execute on the Vera C. Rubin Telescope, and will observe roughly 6 million minor bodies, up to 400 times each, with observations spanning six optical filters: u, g, r, i, z, Y \citep{Ivezic2019}. The image and database outputs of the LSST will be in the petabytes. 

The pipeline that will be used to search for moving bodies in the LSST is largely a classical pipeline, and is developed on top of the decades of experience the ground-based community has in searching for minor bodies using non-ML based techniques \citep{Heinze2022}. The pipeline will search for objects detectable in single visits of 2 back-to-back 15~s exposures. The anticipated search depth from this pipeline is $m_r \sim 24.0$. Due to the short $15$~s exposures, all but the closest near-Earth Objects (NEAs) will appear as point sources in the observations made with Rubin. 

While methods to detect trailed sources might be somewhat fruitful for NEA science, the search for minor bodies in LSST observations stand to benefit most from shift'n'stack efforts. The expected typical visit cadence a minor body will receive is two visits in two different filters in one night, spanning $\sim33$~minutes in a night, though some small fraction $(\sim10\%)$, of objects will receive a 3rd or even a 4th visit.\footnote{Some details of the survey cadence are available at \url{http://survey-strategy.lsst.io/}} Objects will receive another pair of visits again on a few day timescale. Across the annual $\sim2$~month observability window, an object should receive $\sim40$ visits, roughly half of which will be in the g, r, and i filters.  Of the filters offered by Rubin, these are where most minor bodies are brightest. If imagery across a lunation could be stacked, the search depth would be increased by as much as $\sim1.5$~mags, which would increase the discovery sample by nearly a factor of 3. Even across just one week of observing, shift'n'stack would net a $\sim70\%$ increase in the sample of discovered minor bodies. Needless to say, the potential is enormous. 

Across two months, the trajectories of all minor bodies are highly non-linear -- this is true even for KBOs after as little as $4$ days at near-opposition geometries. As such, non-linear shift and stack will be required to maximize the potential search depth available to Rubin observations. The compute complexity of the non-linear shift'n'stack scales with the time baseline of observations, $t$, as $\mathcal{O}(t^4)$ as the time baseline determines the number of unique trajectories that need to be stacked. Even across a lunation, the non-linear shift'n'stack is impossibly intractable without a fundamental algorithm shift. ML may hold the potential for that algorithm shift. 

One possible approach to implement a search across multiple images is to develop a deep CNN to take as input multi-layer consecutive images of the same field, and train it to recognize sources with brightness below the single-image noise floor. This approach seems plausible through the use of artificial source implantation within the LSST imagery themselves. One of the weaknesses of this approach though is in the fixed size of input imagery, thereby fixing the number of images that can be searched. Not all sources will experience the same observing pattern, and so multiple networks will need to be trained to manage the range of visit circumstances that occur in the dataset.

Another possible approach is through the use of techniques that capture the latent structure of the scene. These techniques include U-nets and image transformers. One can envision an approach that merges recurrent network structures with transformers, whereby a transformer could be trained to produce a weight map marking possible locations of sub-noise moving objects, and a recurrent network that cycles over a temporal sequence of transformer outputs. Such a network would not be limited to an image sequence of fixed length. Regardless of the winning solution, it is inevitable that the approach to non-linear shift'n'stack will involve engineering of advanced networks well beyond the basic styles of CNN that are commonly in use nowadays. If a method is found however, the potential scientific output for Solar System science is astronomical. 

\section{Acknowledgements}
I would like to thank Sebastien Fabbro, Hossen Teimoorinia, and Preeti Cowan for their helpful suggestions regarding overcoming training issues and improvements to the neural networks I have worked with. I would also like to thanks JJ Kavelaars and John Spencer for providing the encouragement needed to learn machine learning techniques in the first place. Finally, I would like to thank Sam Lawler for helping execute the CLASSY program the data products of which many of these results are derived.

Based in part on observations obtained with MegaPrime/MegaCam, a joint project of CFHT and CEA/DAPNIA, at the Canada-France-Hawaii Telescope (CFHT) which is operated by the National Research Council (NRC) of Canada, the Institut National des Science de l'Univers of the Centre National de la Recherche Scientifique (CNRS) of France, and the University of Hawaii. The observations at the Canada-France-Hawaii Telescope were performed with care and respect from the summit of Maunakea which is a significant cultural and historic site. 

\section{Code availability}
\label{sec:code}

The trained flux regressor models discussed above, and a Jupyter notebook demonstrating their training and usage are available at \url{https://solar-system-ml.github.io/book/chapter9/cnn_mags_production/} with associated documentation available at \url{https://solar-system-ml.github.io/book}.\\

\section{References}


\clearpage
\Backmatter
%
%
%
%
%
%
%
%
%
\bibliographystyle{elsarticle-harv}

%
%
%
%
%
%
%
%
%
%
%
%
%
%
%
%
%

\end{document}